\documentclass[aps,prd,preprintnumbers,superscriptaddress,nofootinbib,notitlepage,floatfix,10pt]{revtex4-2}
\usepackage[pdftex]{graphicx}
\usepackage{bm,latexsym,amsmath,amssymb,amsfonts,mathrsfs}
\usepackage{physics}
\usepackage{picture}
\usepackage{here}
\usepackage{color}
\allowdisplaybreaks[1]
\usepackage[pdftex,colorlinks=true,linkcolor=blue,citecolor=cyan,backref=page]{hyperref}
\newcommand*{\D}{\mathrm{d}}

\newcommand{\cd}{\nabla}
\newcommand{\mrm}[1]{\mathrm{#1}}

\newcommand{\mcal}[1]{\mathcal{#1}}

\def\om{\Omega_{\mathrm{m}}}
\def\omp{\Omega_{\mathrm{m}0}}

\begin{document}
\title{Spherical collapse in DHOST theories and EFT of dark energy}
%
\author{Toshiki~Takadera}
\email[Email: ]{t\_takadera@rikkyo.ac.jp}
\affiliation{Department of Physics, Rikkyo University, Toshima, Tokyo 171-8501, Japan}
\author{Takashi~Hiramatsu}
\email[Email: ]{hiramatsu.takashi@nihon-u.ac.jp}
\affiliation{Department of Physics, College of Science and Technology, Nihon University, 1-8-14 Kanda-Surugadai, Chiyoda-ku, Tokyo 101-8308, Japan}
\affiliation{Department of Physics, Rikkyo University, Toshima, Tokyo 171-8501, Japan}
\author{Tsutomu~Kobayashi}
\email[Email: ]{tsutomu@rikkyo.ac.jp}
\affiliation{Department of Physics, Rikkyo University, Toshima, Tokyo 171-8501, Japan}
%
\begin{abstract}
We study the nonlinear evolution of matter overdensities using the spherical collapse model in degenerate higher-order scalar-tensor (DHOST) theories beyond Horndeski, employing the effective field theory (EFT) of dark energy approach.
We investigate the impact of the EFT parameters characterising DHOST theories on the formation of large-scale structure.
We identify the parameter space in which the collapse of the spherical overdensity is prevented by the scalar field turning imaginary at some moment, which allows us to place constraints on the model parameters.
We show how the collapse time and the critical density contrast depend on the EFT parameters.
To assess the observational implications, we compute the halo mass function using the Press-Schechter formalism.
We find that the number density of halos is suppressed compared to the $\Lambda$CDM model due to ``beyond Horndeski'' effects, upon imposing the stability of linear perturbations.
\end{abstract}
\preprint{RUP-25-9}
\maketitle

\section{Introduction}\label{sec:intro}

The observation of type Ia supernovae revealed that the Universe is undergoing a period of accelerated expansion~\cite{SupernovaCosmologyProject:1998vns,SupernovaSearchTeam:1998fmf}.
The most widely accepted model to explain this accelerated expansion is the $\Lambda$CDM model, in which the cosmological constant $\Lambda$ drives the acceleration of the Universe.
This model has achieved remarkable success in accounting for a wide range of cosmological observations with a small number of cosmological parameters.
However, it faces a major challenge: the value of $\Lambda$ must be extremely fine-tuned to be consistent with the current accelerated expansion.
Moreover, the $\Lambda$CDM model exhibits mild but persistent tensions with certain observational data.
These issues motivate us to explore alternative explanations of cosmic acceleration.
One such possibility is to modify gravity on cosmological scales.

Scalar-tensor theories have been the most extensively studied among various modified gravity models, as they represent the simplest class, and more intricate models are often described effectively as scalar-tensor theories.
Such theories extend general relativity by introducing an additional scalar degree of freedom on top of the two tensorial gravitational-wave degrees of freedom.
Degenerate higher-order scalar-tensor (DHOST) theories~\cite{Langlois:2015cwa,Crisostomi:2016czh,BenAchour:2016fzp} offer a general and comprehensive framework of modified gravity with $(2+1)$ degrees of freedom, encompassing the Horndeski family~\cite{Horndeski:1974wa}, i.e., the most general family of scalar-tensor theories with second-order field equations, as a subset.
See Refs.~\cite{Langlois:2018dxi,Kobayashi:2019hrl} for a review.
Over the past decade, there has been significant progress in various aspects of DHOST theories, including cosmological solutions~\cite{Crisostomi:2018bsp}, the linear growth of structure~\cite{Hirano:2019nkz}, constraints from the CMB~\cite{Hiramatsu:2020fcd,Hiramatsu:2022fgn}, primordial gravitational waves from inflation~\cite{Brax:2025osk}, screening mechanisms~\cite{Kobayashi:2014ida,Langlois:2017dyl,Crisostomi:2017lbg,Dima:2017pwp,Hirano:2019scf}, and the relativistic stellar structure~\cite{Babichev:2016jom,Sakstein:2016oel,Kobayashi:2018xvr,Kobayashi:2025bdh}.

In this paper, we explore potential tests of DHOST theories.
In the Galileon/Horndeski class of scalar-tensor theories, the Vainshtein mechanism operates due to the nonlinear derivative interactions of the scalar degree of freedom, leading to the recovery of general relativity in the vicinity of a source~\cite{Kimura:2011dc,Narikawa:2013pjr,Koyama:2013paa}.
However, DHOST theories beyond the Horndeski class exhibit a partial breaking of the screening mechanism inside material bodies~\cite{Kobayashi:2014ida,Langlois:2017dyl,Crisostomi:2017lbg,Dima:2017pwp,Hirano:2019scf}.
The breaking of Vainshtein screening has been studied so far assuming a static and spherically symmetric setup.
Cosmological density perturbations offer us a time-dependent setup, but the effects of Vainshtein screening and its breakdown are not apparent at the level of linear perturbations.
The present paper focuses on large-scale structure of the Universe in the context of DHOST theories, where both time dependence and nonlinearity play an essential role.


Large-scale structure of the Universe is one of the most promising observational probes for testing gravity on cosmological scales (see, e.g., \cite{Arai:2022zzz} for a review).
To confront gravity theories with large-scale structure observations, it is crucial to understand the nonlinear growth of matter overdensities in modified gravity.
The spherical collapse model provides a simple and semi-analytical framework to describe this nonlinear evolution.
It has previously been applied to the Galileon/Horndeski class of theories as well as other classes of modified gravity theories.
In Ref.~\cite{Bellini:2012qn}, the authors studied the spherical collapse model in covariant Galileon gravity~\cite{Deffayet:2009mn,Deffayet:2009wt} and investigated the impact of Galileon terms on the critical density contrast.
Reference~\cite{Barreira:2013xea} studied the halo mass function in covariant Galileon gravity using the excursion set formalism.
A similar analysis was carried out in Ref.~\cite{Frusciante:2020zfs} in the context of Galileon ghost condensate model~\cite{Deffayet:2010qz,Kase:2018iwp}.
The authors of Ref.~\cite{Albuquerque:2024hwv} considered the spherical collapse model in shift-symmetric Galileon theories using the effective field theory (EFT) of dark energy approach.

In this paper, we extend the analysis of Ref.~\cite{Albuquerque:2024hwv} by applying the spherical collapse model to DHOST theories within the EFT of dark energy framework.
A key advantage of the EFT approach is that it allows one to analyze the theory without specifying the detailed forms of the free functions in the action, while one needs to assume the background evolution and the time dependence of the EFT coefficients.
We first formulate the spherical collapse model in terms of EFT coefficients characterizing DHOST theories.
We then compute the nonlinear evolution of matter overdensities and investigate the potential impact of DHOST theories on the large-scale structure. While doing so, we consider two different background models and clarify how the assumption on the background model affects the final results.
Additionally, we compute the halo mass function using the Press-Schechter formalism.


This paper is organized as follows.
In the next section, we briefly review the EFT of dark energy and its relation to DHOST theories, keeping the nonlinear derivative interactions relevant to the Vainshtein screening mechanism.
In Sec.~\ref{sec:Basic_equations}, we present the basic equations for the spherical collapse model in the EFT of dark energy and DHOST theories, together with the background cosmological models we assume in this paper.
We then analyze the evolution of spherical overdensities in Sec.~\ref{sec:Evolution}.
In Sec.~\ref{sec:Mass_function}, we compute the halo mass function using the Press-Schechter formalism.
Finally, we draw our conclusions in Sec.~\ref{sec:conclusions}.

\section{EFT of dark energy and scalar-tensor theories}
\label{sec:EFT}

For the analysis of spherical collapse,
we use the EFT of dark energy expressed in terms of the so-called $\alpha$-basis,
which can be derived directly from the action of DHOST theories by a perturbative expansion
around a cosmological background.
For our purpose, we make the quasi-static approximation (or the subhorizon approximation)
and keep the nonlinear derivative interaction terms relevant to the Vainshtein mechanism,
going beyond linear cosmology.
The effective action for the metric in the Newtonian gauge,
\begin{align}
    \D s^2=-[1+2\Phi(t,\Vec{x})]\D t^2+a^2(t)
    [1-2\Psi(t,\Vec{x})]\D\Vec{x}^2,
\end{align}
and the scalar field, $\phi=t+\pi(t,\Vec{x})$,
is given by~\cite{Dima:2017pwp} (see also~\cite{Crisostomi:2017lbg,Langlois:2017dyl})
\begin{align}
    S=\int \D t\D^3x&\biggl\{\frac{M^2a}{2}\biggl[\left(c_1\Phi+c_2\Psi+c_3\pi\right)\cd^2\pi+c_4\Psi\cd^2\Phi+c_5\Psi\cd^2\Psi+c_6\Phi\cd^2\Phi+\left(c_7\dot{\Psi}+c_8\dot{\Phi}+c_9\ddot{\pi}\right)\cd^2\pi
    \notag \\
  &+\frac{b_1}{a^2}\mcal{L}^{\mrm{Gal}}_3+\frac{1}{a^2}(b_2\Phi+b_3\Psi)\mcal{E}^{\mrm{Gal}}_3+\frac{1}{a^2}(b_4\cd_i\Psi+b_5\cd_i\Phi+b_6\cd_i\dot{\pi})\cd_j\pi\cd_i\cd_j\pi
  \notag \\
  &+\frac{1}{a^4}\left(d_1\mcal{L}^{\mrm{Gal}}_4+d_4\cd_i\pi\cd_j\pi\cd_i\cd_k\pi\cd_j\cd_k\pi\right)\biggr]
  -a^3\Phi\delta\rho\biggr\},
    \label{actionEFT}
\end{align}
where a dot stands for differentiation with respect to $t$,
$\nabla_i$ is the spatial derivative, and we defined
\begin{align}
    \mcal{L}^{\mrm{Gal}}_3&:=\frac{1}{2}(\cd\pi)^2\left(\cd^2\pi\right),\\
    \mcal{E}^{\mrm{Gal}}_3&:=\left(\cd^2\pi\right)^2-\cd_i\cd_j\pi\cd_i\cd_j\pi,\\
    \mcal{L}^{\mrm{Gal}}_4&:=\frac{1}{2}(\cd\pi)^2\mcal{E}^{\mrm{Gal}}_3.
\end{align}
We also included the (nonrelativistic) matter overdensity, $\delta\rho=\rho(t,\Vec{x})-\bar\rho(t)=\bar\rho\cdot\delta(t,\Vec{x})$, coupled minimally to gravity, with $\bar\rho(t)$ being the mean matter energy density.
The coefficients of the above action are given in terms of the widely used EFT parameters in the $\alpha$-basis by
\begin{align}
    c_1&=-4H\alpha_B+H[4\alpha_H-2\beta_3(1+\alpha_M)]-2\dot{\beta}_3,\quad c_2=4H(\alpha_M-\alpha_T)+4[H\alpha_H(1+\alpha_M)+\dot{\alpha}_H],\notag \\
    c_3&=-\frac{\bar{\rho}}{M^2}-2H\dot{\alpha}_B-2\dot{H}(1+\alpha_B)+2H^2[\alpha_M-\alpha_B(1+\alpha_M)-\alpha_T] \notag \\
    &\quad 
    +\frac{1}{2}\Big\{H\big[4\dot{\alpha}_H-2(1+\alpha_M)\dot{\beta}_3-\beta_3\dot{\alpha}_M\big]-\ddot{\beta}_3\Big\}\notag \\
    &\quad 
    +\frac{1}{2}\Big\{-H^2(1+\alpha_M)[-4\alpha_H+\beta_3(1+\alpha_M)]+4\dot{H}\alpha_H-\dot{H}(1+\alpha_M)\beta_3\Big\},\notag \\
    c_4&=4(1+\alpha_H),
    \quad 
    c_5=-2(1+\alpha_T),
    \quad 
    c_6=-\beta_3,
    \quad 
    c_7=4\alpha_H,\quad c_8=-2(2\beta_1+\beta_3),\quad 
    c_9=4\beta_1+\beta_3,\notag\\
    b_1&=H[4\alpha_B-\alpha_V(1-\alpha_M)-2\alpha_M+3\alpha_T]-H[8\beta_1\alpha_M+\alpha_H(3+\alpha_M)]+\dot{\alpha}_V-\dot{\alpha}_H-8\dot{\beta}_1,\notag \\
    b_2&=\alpha_V-\alpha_H-4\beta_1,\quad b_3=\alpha_T,\quad b_4=-4\alpha_H,\quad b_5=2(2\beta_1+\beta_3),
    \quad b_6=-2(4\beta_1+\beta_3),\notag \\
    d_1&=-(\alpha_T+\alpha_V-\alpha_H-4\beta_1),\quad d_2=4\beta_1+\beta_3.
\end{align}
In the so-called class-Ia degenerate theories,
the following degeneracy condition is satisfied:
\begin{align}
\beta_3=-2\beta_1\left[2(1+\alpha_H)+\beta_1(1+\alpha_T)\right].
\end{align}
Formally, there is another degeneracy condition that must be satisfied.
However, it is irrelevant in the present quasi-static setup because it is used to remove higher time derivatives.
The explicit expressions for the $\alpha$ parameters in terms of the functions in the DHOST action are presented in Appendix \ref{app:EFT}.
It should be noted here that the above effective action can be used even for large overdensities, $\delta\gg1$, as long as the gravitational potentials remain small, $|\Phi|,|\Psi|\ll1$.


As the entire parameter space is huge, we will focus on a smaller subset by imposing some reasonable assumptions in the following analysis.
First, we assume that the effective Planck mass, $M$, is constant, i.e.,
\begin{align}
    \alpha_M:=\frac{1}{H}\frac{\D\ln M^2}{\D t}=0
\end{align}
for simplicity.
We are interested in the subset of DHOST theories avoiding constraints on the propagation of gravitational waves.
Specifically, the simultaneous detection of gravitational waves (GW170817) and electromagnetic signals (GRB170817A) placed a tight constraint on the deviation of the propagation speed $c_{\mrm{GW}}$ of gravitational waves from that of light~\cite{LIGOScientific:2017zic}.
The condition $c_{\mrm{GW}}=1$ translates to~\cite{Creminelli:2017sry,Baker:2017hug,Sakstein:2017xjx,Ezquiaga:2017ekz}
\begin{align}
  \alpha_T=\alpha_V+\alpha_H=0.
  \label{speed}
\end{align}
We impose this condition throughout the paper.
Furthermore, it has been argued that
\begin{align}
  \alpha_H+2\beta_1=0
  \label{decay}
\end{align}
is required in order for gravitons not to decay into dark energy while they propagate~\cite{Creminelli:2018xsv}.
This condition crucially changes the structure of the equation for $\pi$ and thereby affects how the screening mechanism operates~\cite{Hirano:2019scf,Crisostomi:2019yfo}.
For this reason, we investigate two cases: one where the condition~\eqref{decay} is imposed and one where it is not.
Aside from $M$, the independent parameters are therefore $(\alpha_B,\beta_1)$ in the former case and $(\alpha_B,\alpha_H,\beta_1)$ in the latter case.


\section{Basic equations for spherical collapse in DHOST theory and EFT of dark energy}
\label{sec:Basic_equations}

\subsection{Background model}

In DHOST theories, even the analysis of the evolution of the homogeneous background is complicated due to the (apparent) higher-derivative nature of the field equations~\cite{Crisostomi:2018bsp}.
We therefore assume the background evolution and use the EFT of dark energy/modified gravity to address the evolution of a spherical overdensity on the given background.
In this paper, we simply assume that the background evolution is identical to that of the $\Lambda$CDM model after matter domination.
The Hubble parameter $H=H(t)$ thus obeys
\begin{align}
  \label{Heq}
    \frac{H^2}{H_0^2}=\frac{\omp}{a^3}+1-\omp,
\end{align}
where $H_0$ is the present value of the Hubble parameter and $\omp$ is the matter density parameter. Explicitly, we have 
\begin{align}
    a^3(t)&=\frac{\omp}{1-\omp}\sinh^2 \left[
    \frac{3}{2}(1-\omp)^{1/2}H_0 t
    \right],
    \\ 
    H(t)&=H_0(1-\omp)^{1/2}\coth\left[
    \frac{3}{2}(1-\omp)^{1/2}H_0 t
    \right],
\end{align}
where the scale factor is normalized so that $a=1$ at the present time.
Let us define the time-dependent matter density parameter as 
\begin{align}
  \label{defom}
  \om(t):=\frac{8\pi G_{\mrm{cos}}\bar{\rho}(t)}{3H^2(t)},
\end{align}
where the ``cosmological'' gravitational constant $G_{\mrm{cos}}$ is to be specified below. This quantity can also be expressed in terms of $\omp$ as 
\begin{align}
  \label{omega}
    \om(t)=\frac{H_0^2\omp}{a^3H^2}=\frac{\omp}{\omp+a^3(1-\omp)}.
\end{align}

We also need to specify the time evolution of the $\alpha$ functions.
In this paper, we assume that the time dependence of
$\alpha_i=\{\alpha_B,\alpha_H,\beta_1\}$ is given by
\begin{align}
  \alpha_i(t)= 
  \alpha_{i0}\left[\frac{1-\om(t)}{1-\omp}\right]
  =\alpha_{i0}\left(\frac{H_0}{H}\right)^2,
\end{align}
which is often assumed in the literature.
Here, $\alpha_{i0}$ is the present value of the corresponding $\alpha$ function.
The above assumptions are enough for us to specify the time dependence of the EFT coefficients except for $c_3$.



Let us now come back to the cosmological gravitational constant $G_\mrm{cos}$, which we write
\begin{align}
    G_{\mrm{cos}}=\frac{1}{8\pi M^2\gamma_0},
\end{align}
where $\gamma_0$ is a constant.
Equation~\eqref{defom} then reduces to
\begin{align}
    \bar\rho(t)=3M^2\gamma_0H^2\om(t).
    \label{eq:background_density}
\end{align}
Using this, one can remove $\bar\rho$ that appears in the fluid equations and the EFT coefficient $c_3$.
We consider two different possibilities for the value of $\gamma_0$.
In the first case, we assume that $G_{\mrm{cos}}$ is given by the Newton constant $G_{\mrm{N}}$ measured at the present time, as it is so in the $\Lambda$CDM model in general relativity.
In DHOST theories and the EFT of dark energy,
$G_{\mrm{N}}$ is related to the effective Planck mass, $M$, and the $\alpha$ functions as~\cite{Dima:2017pwp}
\begin{align}
  G_{\mrm{N}}=\frac{1}{8\pi M^2(1-\alpha_H-3\beta_1)}.
\end{align}
Using the values measured at the present time, we set
\begin{align}
  \gamma_0=1-\alpha_{H0}-3\beta_{10}.\label{Gcos1}
\end{align}
In the second case, we simply assume that 
\begin{align}
    \gamma_0=1,\label{Gcos2}
\end{align}
which is often assumed in the literature.
Since $G_{\mrm{N}}= 1/8\pi M^2$ at the beginning of the matter-dominated stage, the latter case amounts to setting $G_{\mrm{cos}}$ equal to $G_{\mrm{N}}$ measured at an early time.

\subsection{The fluid equations}

Since matter is assumed to be minimally coupled to gravity, it obeys the standard continuity and the Euler equations.
The matter overdensity $\delta(t,\Vec{x})$ and the velocity field $\Vec{v}(t,\Vec{x})$ are therefore governed by
\begin{align}
    \dot \delta + \frac{1}{a}\Vec{\nabla}\cdot\left[(1+\delta)\Vec{v}\right]
    &=0,\label{conEq}
    \\ 
    \dot{\Vec{v}}+H\Vec{v}+\frac{1}{a}(\Vec{v}\cdot\Vec{\nabla})\Vec{v}
    &=-\frac{1}{a}\Vec{\nabla}\Phi.\label{EulerEq}
\end{align}
We consider a spherical top-hat overdensity,
\begin{align}
    \delta(t,\Vec{x})=\delta(t)\quad (r=|\Vec{x}|\le r_*),
    \notag
\end{align}
where $r_*$ is the comoving radius of the overdense region.
The velocity field for the spherical top-hat overdensity is given by 
\begin{align}
    \Vec{v}(t,\Vec{x})=\frac{1}{3}\theta(t)\Vec{x},
\end{align}
where we have $\theta=\Vec{\nabla}\cdot\Vec{v}$.
Substituting these to Eqs.~\eqref{conEq} and~\eqref{EulerEq},
we obtain
\begin{align}
    \dot\delta+\frac{1}{a}(1+\delta)\theta&=0,
    \\ 
    \dot\theta+H\theta+\frac{\theta^2}{3a}&=-\frac{1}{a}\nabla^2\Phi,
\end{align}
which are combined to give 
\begin{align}
    \ddot\delta+2H\dot\delta-\frac{4}{3}\frac{\dot\delta^2}{1+\delta} 
    =(1+\delta)\frac{\nabla^2\Phi}{a^2}.
\end{align}
This equation implies that $\nabla^2\Phi=r^{-2}(r^2\Phi')'$ is independent of $r$.
The gravitational potential that satisfies this requirement and is regular at $r=0$ is of the form $\Phi=\mcal{Y}(t)r^2$, and hence 
$\nabla^2\Phi=3\Phi'/r=6\mcal{Y}(t)$, where 
the prime denotes differentiation with respect to $r$.
The evolution equation for $\delta$ is thus given by 
\begin{align}
    \ddot\delta+2H\dot\delta-\frac{4}{3}\frac{\dot\delta^2}{1+\delta} 
    =3(1+\delta)\frac{\Phi'}{a^2r},\label{deltaEq}
\end{align}
where $\Phi'/r$ is dependent only on $t$.

\subsection{Gravitational field equations}

Now we return to the action~\eqref{actionEFT} and derive the gravitational field equations.
By relating the gravitational potential $\Phi$ with $\delta$ through the modified Poisson equation, one can close the system of equations and solve Eq.~\eqref{deltaEq}.
To do so, it is convenient to introduce new dimensionless variables
\begin{align}
  x(t):=\frac{\pi'}{H_0a^2r},\qquad 
  y(t):=\frac{\Phi'}{H_0^2a^2r},\qquad 
  z(t):=\frac{\Psi'}{H_0^2a^2r}.
\end{align}
Note that these quantities depend only on $t$ for a top-hat overdensity,
as will be seen directly from the field equations.

Varying the action~\eqref{actionEFT},
it is straightforward to obtain the equations of motion for $\Phi$, $\Psi$, and $\pi$.
The resultant equations of motion can be integrated once with respect to $r$,
leading to the equations schematically written as
\begin{align}
  &\mcal{F}_0(x,\dot{x},y,\dot{y},z,\dot{z})=0,\label{xeqintNL}
  \\
  &y=Y(x,\dot{x},\delta),\label{yeqintNL}
  \\
  &z=Z(x,\dot{x},\delta).\label{zeqintNL}
\end{align}
Here, the equations of motion for $\Phi$ and $\Psi$ were rearranged and transformed into the forms of Eqs.~\eqref{yeqintNL} and~\eqref{zeqintNL}, and integration constants were fixed so that the equations admit the solution $x=y=z=0$ when $\delta=0$.
One can eliminate $y$ and $z$ from Eq.~\eqref{xeqintNL} by using Eqs.~\eqref{yeqintNL} and~\eqref{zeqintNL}, yielding a cubic algebraic equation for $x$:
\begin{align}
  \mcal{F}(x,\delta,\dot{\delta})=\mcal{C}_3x^3+\mcal{C}_2x^2+\mcal{C}_1x+\mcal{C}_0=0.\label{xeq}
\end{align}
The coefficients are written in terms of the background quantities ($H$ and $\Omega_\mrm{m}$), the $\alpha$ functions, and $\delta$.
For instance,
\begin{align}
    \mcal{C}_3=(\alpha_H+2\beta_1)(1-\alpha_H-3\beta_1).
    \label{eq:cf-C3}
\end{align}
The explicit expressions for the other coefficients are given in Appendix~\ref{app:coefficient}.
Note here that upon substituting Eqs.~\eqref{yeqintNL} and~\eqref{zeqintNL} to Eq.~\eqref{xeqintNL}, the time derivatives of $x$ are canceled thanks to the degeneracy condition, and we are left with the algebraic equation~\eqref{xeq}.

We mentioned earlier that the condition $\alpha_H+2\beta_1=0$ crucially changes the structure of the equation for $\pi$.
This point can be seen clearly from Eq.~\eqref{eq:cf-C3}: the cubic equation reduces to the quadratic one when $\alpha_H+2\beta_1=0$.
We will treat this case separately in the following analysis.

The standard way of solving the system is as follows.
Equation~\eqref{xeq} can be solved to give $x$ in terms of $\delta$ and $\dot\delta$, which allows us to express $y$ in terms of $\delta$ and its derivatives through Eq.~\eqref{yeqintNL}.
The result is regarded as the modified Poisson equation, which replaces
the right-hand side of Eq.~\eqref{deltaEq} with $\delta$ and its derivatives.
One can thus obtain a closed evolution equation for $\delta$.

We will take, however, a technically different approach
in this paper.
Differentiating Eq.~\eqref{xeq} with respect to $t$, we obtain the equation containing $x$, $\dot x$, $\delta$, $\dot\delta$, and $\ddot\delta$:
\begin{align}
    \mcal{G}(x,\dot{x},\delta,\dot{\delta},\ddot{\delta}):=
    \frac{\D\mcal{F}}{\D t}=0.
    \label{dx}
\end{align}
The right-hand side of Eq.~\eqref{deltaEq} can be expressed in terms of $x$, $\dot x$, and $\delta$ by the use of Eq.~\eqref{yeqintNL}.
One thus arrives at a system of differential equations that is second order for $\delta$ and first order for $x$, supplemented with the constraint equation~\eqref{xeq}.
First, we set the initial conditions $\delta=\delta_i$ and $\dot\delta=\dot\delta_i$ at some initial moment. Next, the constraint equation~\eqref{xeq} is solved at the initial moment to give the initial condition $x=x_i$.
The set of differential equations~\eqref{deltaEq} and~\eqref{dx} can then be solved to determine $\delta(t)$ and $x(t)$.
When doing this procedure numerically, Eq.~\eqref{xeq} is useful for verifying the accuracy of the numerical calculations at each time step.

To describe the evolution of a spherical top-hat overdensity,
it is often convenient to introduce a variable
\begin{align}\label{defR}
    R(t):=\frac{a(t)r_*}{[1+\delta(t)]^{1/3}}
\end{align}
in place of $\delta$,
which is nothing but the physical radius of the spherical overdense region.

\subsection{Linear regime}

When $x$, $y$, $z$, and $\delta$ are sufficiently small, we can approximate their governing equations by their linearized forms.
The linear density field $\delta_L$ thus obeys
\begin{align}
    \ddot\delta_L+2H\dot\delta_L=3H_0^2y,\label{deltaEqL}
\end{align}
while from the gravitational field equations we have 
\begin{align}
    \mcal{C}_1x+\mcal{C}_0&=0,\label{lin-eq-1}
    \\ 
         (1+\alpha_H+\beta_1)^2y&=
         \frac{H}{H_0}\left[\alpha_B+(\alpha_H+\beta_1)(1+2\alpha_H+2\beta_1)\right]x+\frac{\gamma_0}{2}\frac{H^2}{H_0^2}\om\delta_L-\dot{\beta_1}x
         \notag \\ &\quad 
    +(\alpha_H+\beta_1)(1+\alpha_H+\beta_1)\dot{x}.
    \label{lin-eq-2}
\end{align}
Using Eq.~\eqref{lin-eq-1} one can remove $x$ from Eq.~\eqref{lin-eq-2} to express $y$ in terms of $\delta_L$ and its derivatives.
Plugging this into Eq.~\eqref{deltaEqL}, one can derive the evolution equation for the linear density field $\delta_L$, which will be used later to compute the critical density contrast.

We can use the above equations to set the initial condition for $\delta$ deep in the matter-dominated era.
Since we are assuming that $\alpha_i\propto 1-\om(t)$,
the $\alpha$ functions are well suppressed when $\om\simeq 1$, which leads to
$y\simeq (\gamma_0/2)(H^2/H_0^2)\om\delta_L=(\gamma_0/2H_0^2)(4/9t^2)\delta_L$.
Thus, deep in the matter-dominated era, we obtain
\begin{align}
    \ddot\delta_L+\frac{4}{3t}\dot\delta_L=\frac{2\gamma_0}{3t^2}\delta_L.
    \label{eq:linear}
\end{align}
The growing solution is given by
\begin{align}
     \delta_L \propto a^{n},\qquad n=\frac{1}{4}\left(
    -1+\sqrt{24+\gamma_0}\right).
\end{align}
We use the linearized solution for the initial condition of $\delta$, as will be discussed in the next section.

Equation~\eqref{xeq} has three solutions in general, but only one of them is connected smoothly to the linear solution in the early stage.
Note in passing that the potential difference from the standard linear result in general relativity comes from the fact that the effective gravitational coupling for the linear density field differs from $G_\mrm{cos}$ by the factor of $\gamma_0$.
In the case of $\gamma_0=1$, we have the conventional result $\delta\propto a$.

\section{Evoluiton of spherical overdensities}
\label{sec:Evolution}

\subsection{The case of $\alpha_H+2\beta_1=0$}

In this section, we consider the special case where $\alpha_H+2\beta_1=0$ is satisfied,
with the background model being described by Eq.~\eqref{Gcos1}.
In this case, Eq.~\eqref{xeq} reduces to a quadratic equation, and one of the two roots is connected to the linear regime in the early stage.
The important point is that we need to require that
$\mcal{D}:=\mcal{C}_1^2-4\mcal{C}_2\mcal{C}_0\ge 0$
in order for $x$ to remain real over the entire history of the evolution of $\delta$. We will see in the following that a tiny value of $\beta_{10}$ can easily drive $x$ to be imaginary, leading to tight limits on the allowed value of $\beta_{10}$.

\begin{figure}[tb]
  \begin{minipage}[b]{0.45\linewidth}
    \centering
    \includegraphics[keepaspectratio, scale=0.45]{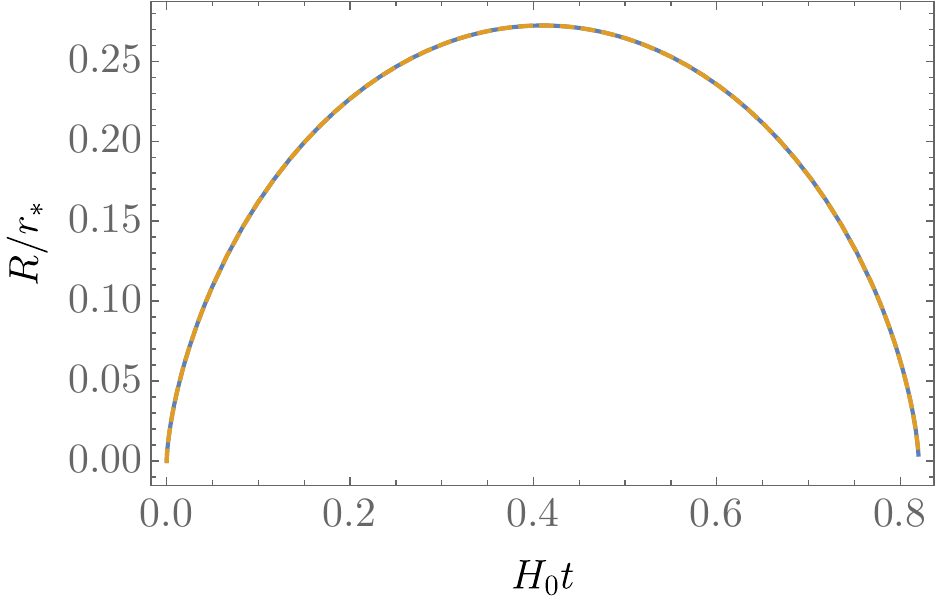}
  \end{minipage} 
  \begin{minipage}[b]{0.45\linewidth}
    \centering
    \includegraphics[keepaspectratio, scale=0.45]{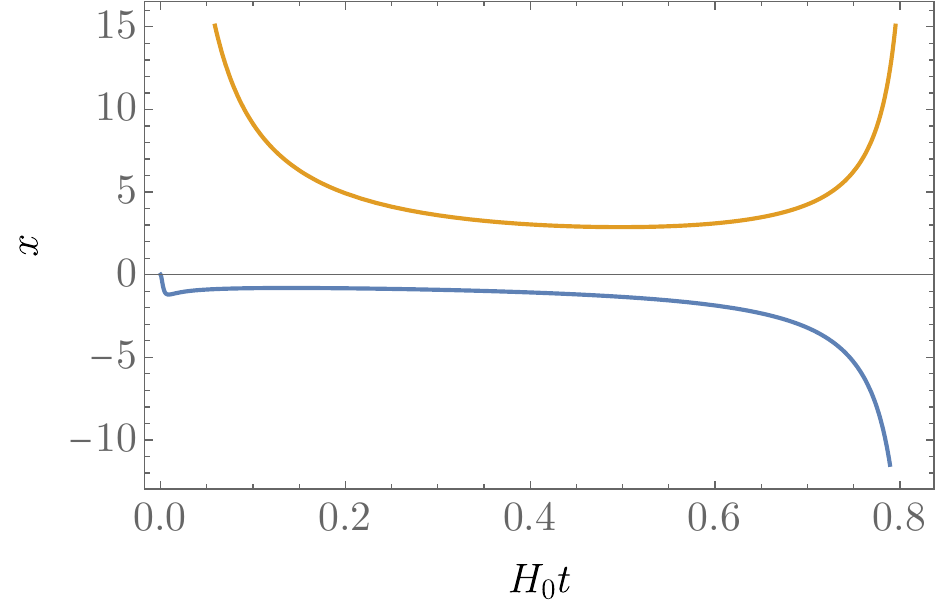}
  \end{minipage} \\
  \caption{Evolution of $R$ (left) and $x$ (right) in the case where real solutions for $x$ continue to exist until the collapse is completed.
  The solid line in the left panel is for $\beta_{10}=10^{-7}$, $\alpha_{B0}=-2\times10^{-3}$, and $A_i=2.30$, while the dashed line in the left panel is for $\beta_{10}=0$, $\alpha_{B0}=-2\times10^{-3}$, and $A_i=2.30$, though the two lines are overlapping, making them indistinguishable.
  The blue line in the right panel is for $\beta_{10}=10^{-7}$, $\alpha_{B0}=-2\times10^{-3}$, and $A_i=2.30$ and starts from the linear regime.
  Another branch that is not connected to the linear solution in the early stage is shown in orange.
  }
 \label{collapse_b_7}
\end{figure}

\begin{figure}[tb]
  \begin{minipage}[b]{0.45\linewidth}
    \centering
    \includegraphics[keepaspectratio, scale=0.45]{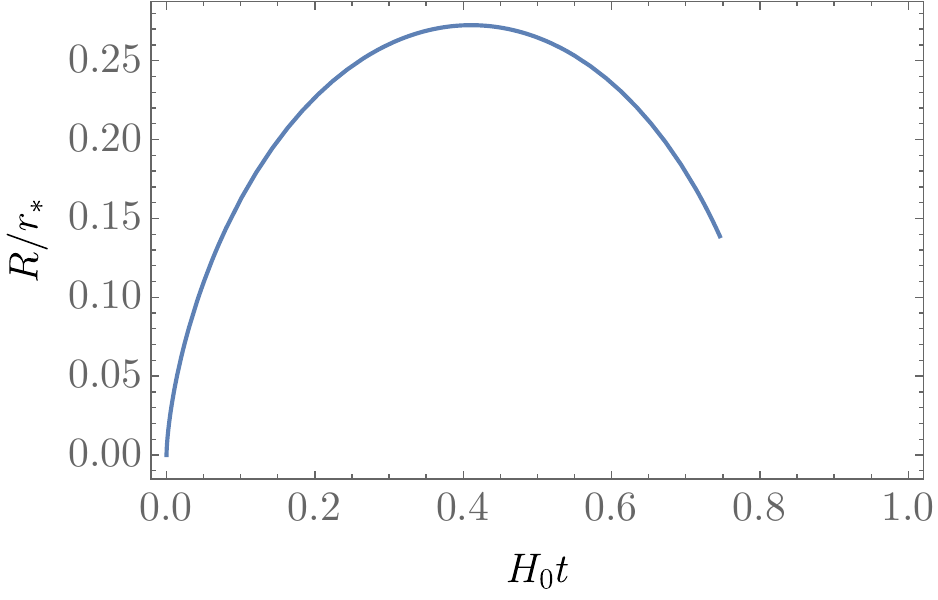}
  \end{minipage} 
  \begin{minipage}[b]{0.45\linewidth}
    \centering
    \includegraphics[keepaspectratio, scale=0.45]{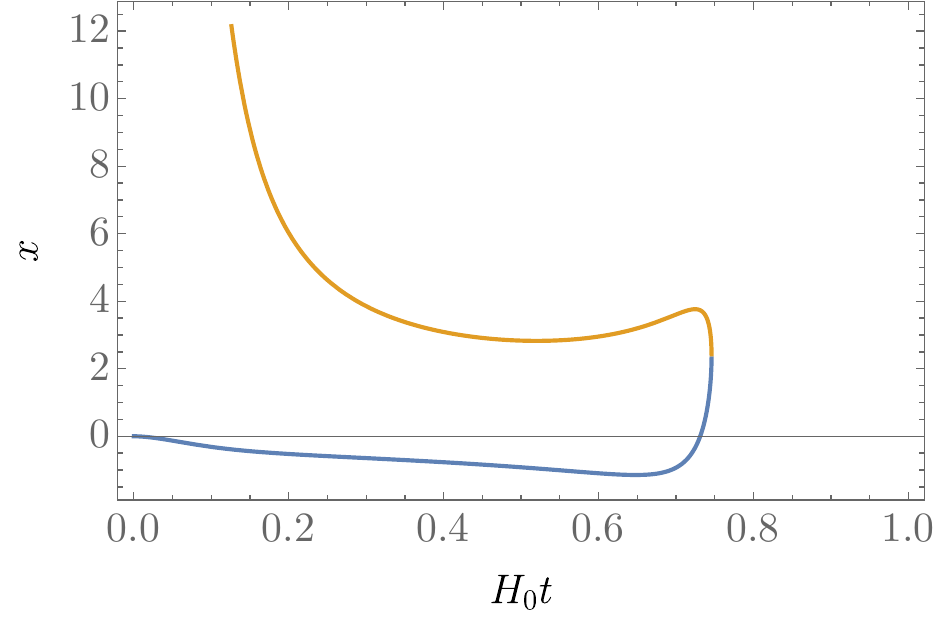}
  \end{minipage} \\
  \caption{Evolution of $R$ (left) and $x$ (right) in the case where real solutions for $x$ cease to exit at some moment. The parameters are given by $\beta_{10}=10^{-4}$, $\alpha_{B0}=-2\times10^{-3}$, and $A_i=2.30$.
  In the right panel, the blue line starts from the linear regime.
  This and another solution (shown in orange) merge at $H_0t\simeq 0.745 $ when $\mcal{D}$ becomes zero.
  }
 \label{collapse_b_4}
\end{figure}

We set the initial condition for $\delta$ as
\begin{align}
    \delta(t_i) =A_i a^n(t_i),
    \quad  
    \dot \delta(t_i) =nH(t_i)\delta (t_i),\label{eq:init-delta}
\end{align}
at $t=t_i:=10^{-8}H_0^{-1}$, where $A_i$ is a constant and $n$ was already introduced above.
We fix $\alpha_{B0}=-2\times 10^{-3}$ and investigate the impacts of $\beta_{10}\,(>0)$ on the evolution of $\delta$.
(See Appendix \ref{app:para-region} for the reason why we exclude negative $\beta_{10}$~\cite{Crisostomi:2019yfo}.)
Although the overdense region collapses for sufficiently small $\beta_{10}$, we find that, for $\beta_{10}$ above a certain value, $\mcal{D}$ vanishes at some moment and the model fails to describe the entire evolution of the collapsing region.
In Fig.~\ref{collapse_b_7}, we present the evolution of $R$ and $x$ for $\beta_{10}=10^{-7}$ and $A_i=2.30$.\footnote{For $A_i=2.30$, the redshift corresponding to the collapse time is given by $z\simeq 0.14$.} In this case, the spherical collapse successfully proceeds, but $\beta_{10}$ is so small that its impacts on the evolution of $R$ are negligible, resulting in the evolution of $\delta$ indistinguishable from that in the cubic Galileon model.
Figure~\ref{collapse_b_4} shows the evolution of $R$ and $x$ for $\beta_{10}=10^{-4}$ and $A_i=2.30$. Even for such a tiny value of $\beta_{10}$,
real solutions for $x$ cease to exist at some moment.
By requiring that real solutions for $x$ exist until the present time with any initial amplitudes, we find the upper limit on $\beta_{10}$ as $\beta_{10}<10^{-7}$ for the fiducial value $\alpha_{B0}=-2\times 10^{-3}$.

Real solutions for $x$ cease to exist when $\mathcal{D}$ becomes zero.
To see how this occurs, let us write $\mathcal{D}$ in the following way:
\begin{align}
\label{D}
    \mathcal{D}&=D_1+D_2\delta+D_3\delta^2-D_4\dot{\delta},
    \\ 
    D_4&=4\beta_1(1-\beta_1)^2\left[\dot\beta_1-H(\alpha_B+\beta_1)\right]\gamma_0 
    \frac{H^2\Omega_{\textrm{m}}}{H_0^3},
\end{align}
where the coefficients $D_i$ ($i=1,2,3$) are also written in terms of the $\alpha$ functions.
As we choose the parameter values so that $x$ is real for a static source, the first three terms in Eq.~\eqref{D} are positive (see Appendix~\ref{app:para-region}).
In contrast, the last term in Eq.~\eqref{D}, which is specific to time-dependent systems in DHOST theories, is negative (i.e., $D_4>0$) for $\beta_1>0$ and $\alpha_B<0$ with $\beta_1<|\alpha_B|$.
For $\beta_1>|\alpha_B|$, $D_4$ is still positive at least in the matter-dominated era.
This is how the last term in Eq.~\eqref{D} can hinder the prolonged evolution of $\delta$.

The above results have been derived for the background model given by Eq.~\eqref{Gcos1}.
However, even if we adopt the different background model with $\gamma_0=1$, the results do not change significantly, and we obtain a comparable constraint on $\beta_{10}$.

\subsection{The case of $\alpha_H+2\beta_1\neq 0$}

\begin{figure}[tb]
\begin{tabular}{cc}
  \begin{minipage}[b]{0.45\linewidth}
    \centering
    \includegraphics[keepaspectratio, scale=0.45]{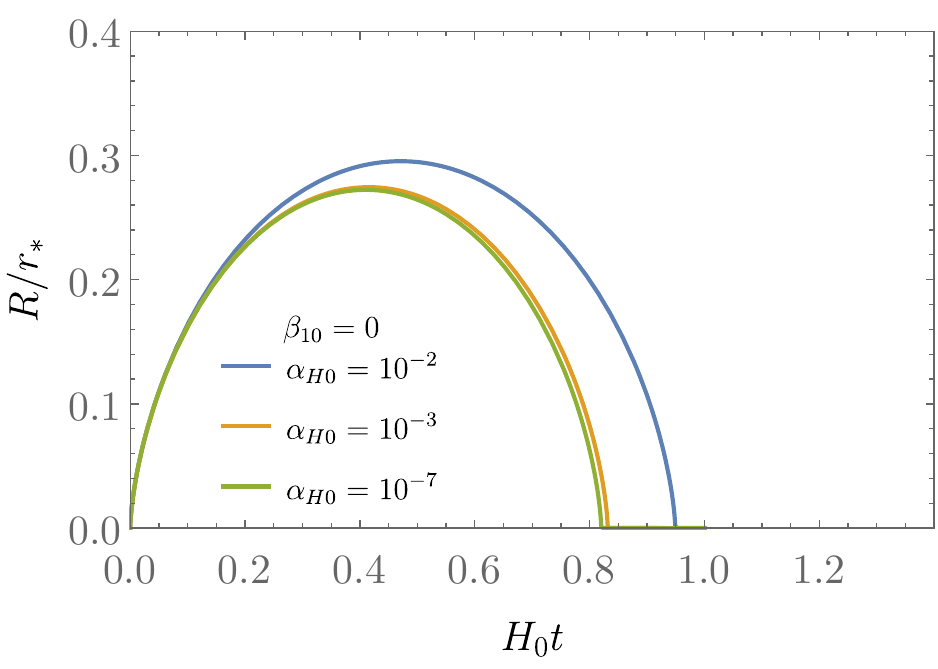}
  \end{minipage} &
  \begin{minipage}[b]{0.45\linewidth}
    \centering
    \includegraphics[keepaspectratio, scale=0.45]{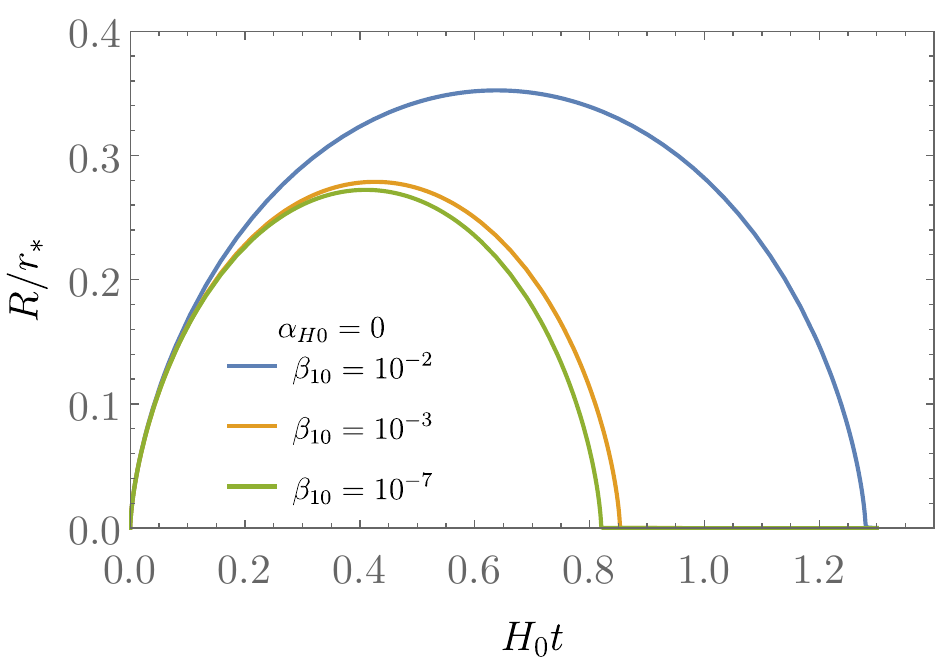}
  \end{minipage}\\
   \begin{minipage}[b]{0.45\linewidth}
    \centering
    \includegraphics[keepaspectratio, scale=0.45]{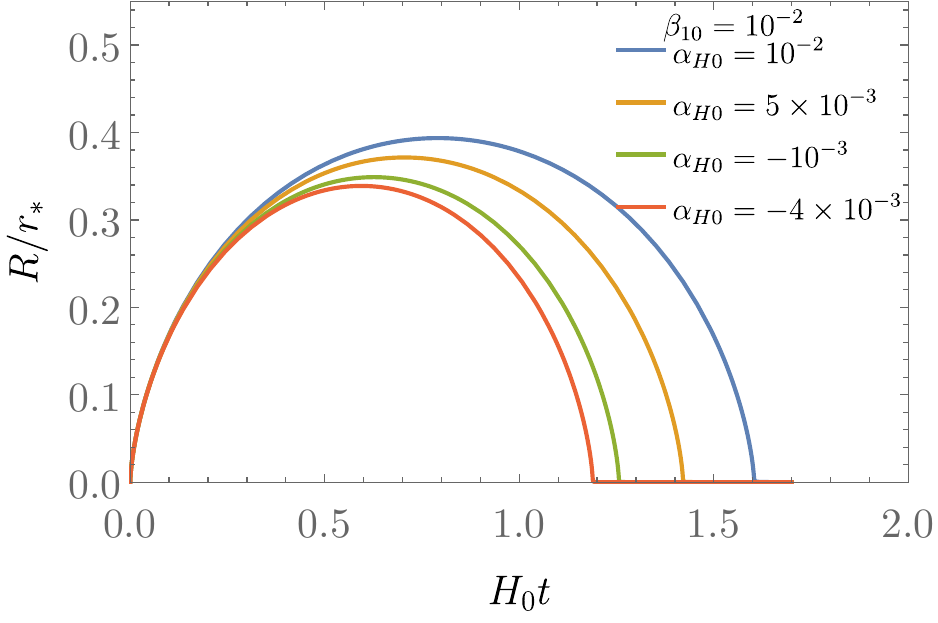}
  \end{minipage} &
  \begin{minipage}[b]{0.45\linewidth}
    \centering
    \includegraphics[keepaspectratio, scale=0.45]{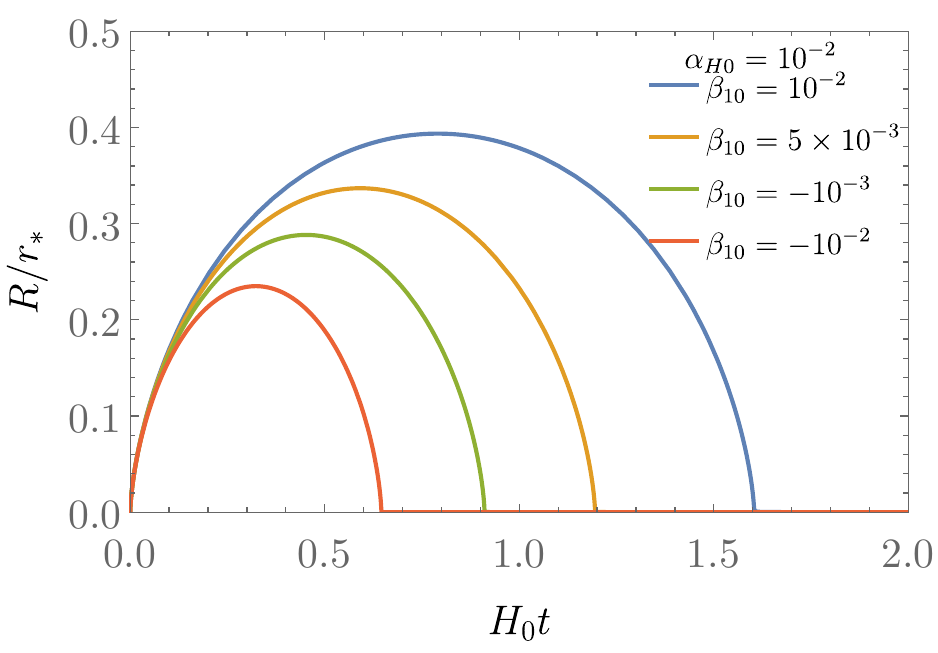}
  \end{minipage}
   \end{tabular}
  \caption{Evolution of $R$ for different $\alpha_{H0}$ and $\beta_{10}$.
  We set $\alpha_{B0}=-2\times10^{-3}$.
  The background model is given by $\gamma_0=1-\alpha_{H0}-3\beta_{10}$.}
 \label{collapse_cubic}
\end{figure}

\begin{figure}[tb]
\begin{tabular}{cc}
  \begin{minipage}[b]{0.45\linewidth}
    \centering
    \includegraphics[keepaspectratio, scale=0.45]{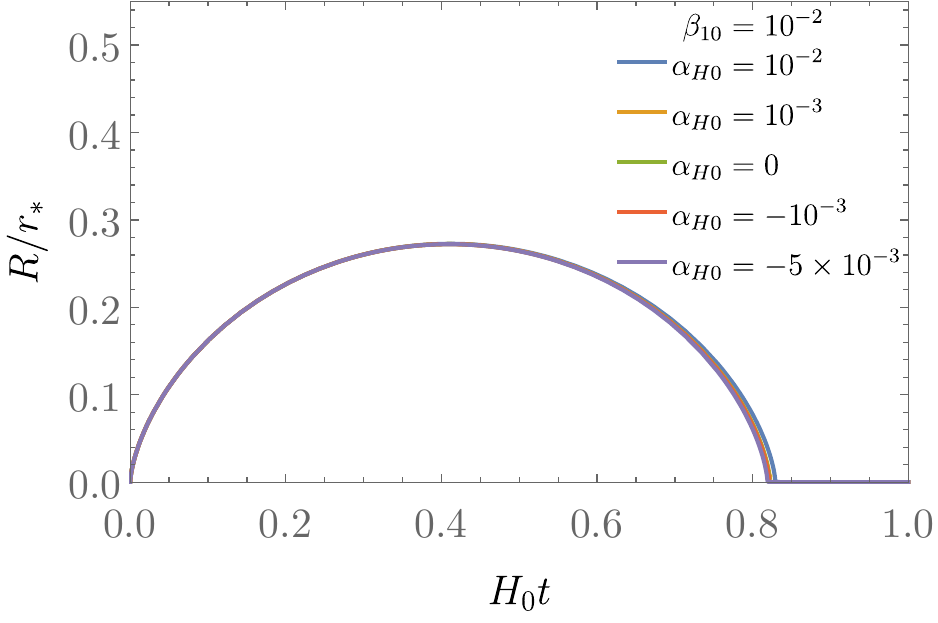}
  \end{minipage} &
  \begin{minipage}[b]{0.45\linewidth}
    \centering
    \includegraphics[keepaspectratio, scale=0.45]{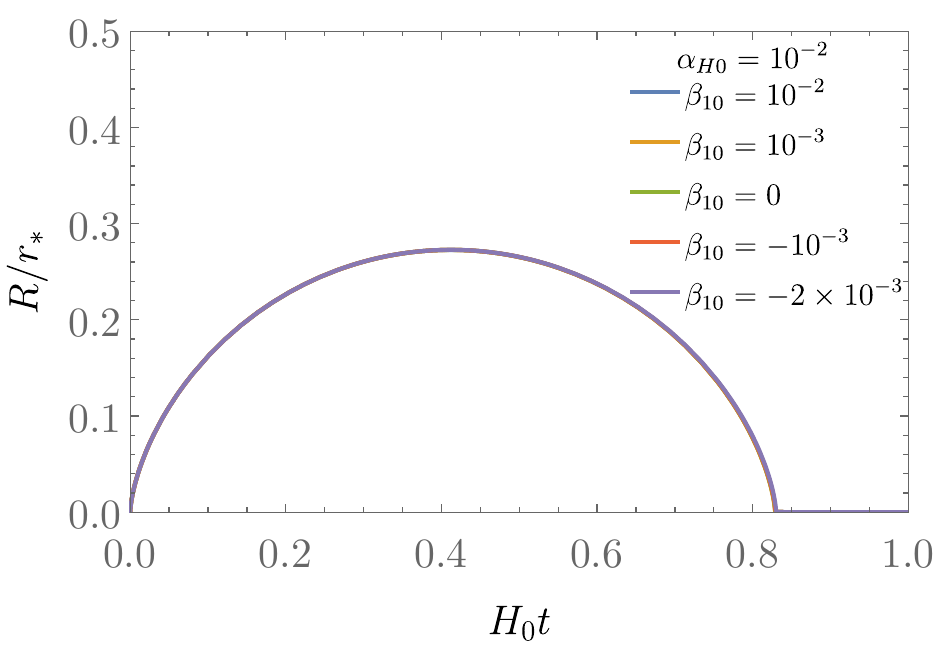}
  \end{minipage}\\
   \begin{minipage}[b]{0.45\linewidth}
    \centering
    \includegraphics[keepaspectratio, scale=0.45]{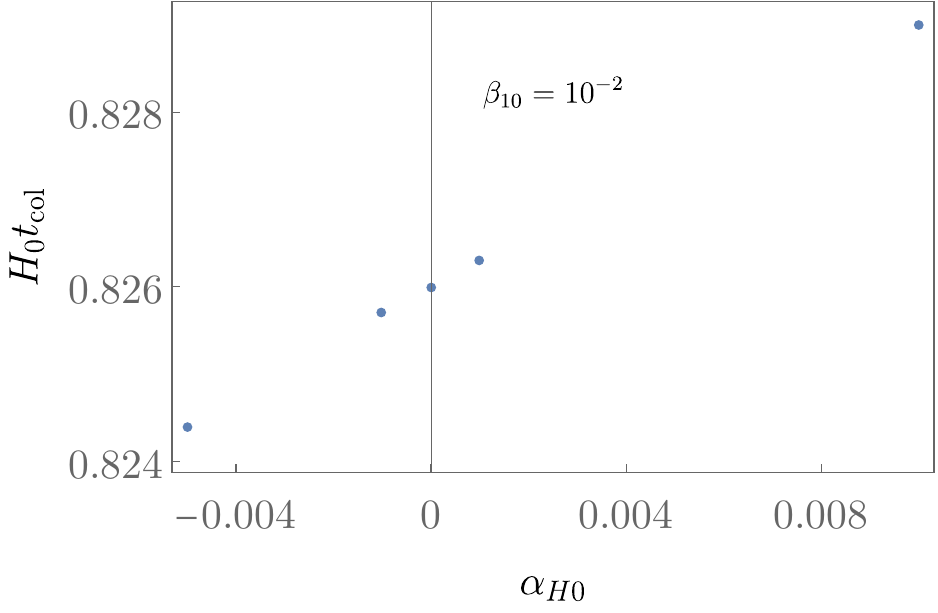}
  \end{minipage} &
  \begin{minipage}[b]{0.45\linewidth}
    \centering
    \includegraphics[keepaspectratio, scale=0.45]{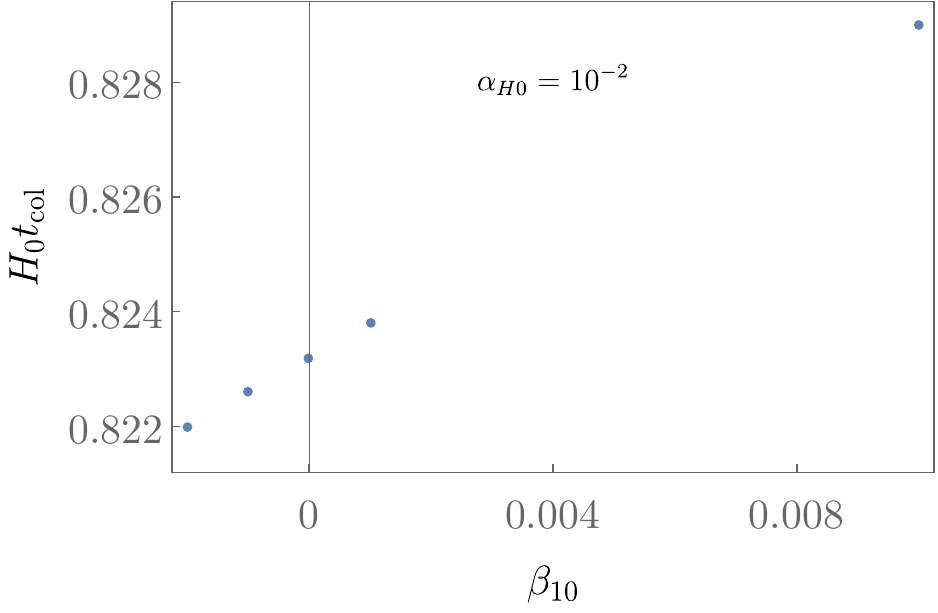}
  \end{minipage}
   \end{tabular}
  \caption{Evolution of $R$ for the background model with $\gamma_0=1$ (top left and top right) and parameter dependence of the collapse time $t_{\mrm{col}}$ (bottom left and bottom right). 
  We set $\alpha_{B0}=-2\times10^{-3}$ and $A_i=2.30$ as the initial amplitude.}
 \label{collapse_cubic2}
\end{figure}

Let us move to the generic case with $\alpha_H+2\beta_1\neq 0$.
Now Eq.~\eqref{xeq} has three roots, one of which starts from the linear regime, $x\ll 1$.
We require that $x^2\propto \delta$ be satisfied for $\delta \gg 1$ so that the Vainshtein mechanism operates outside of an astrophysical body~\cite{Kobayashi:2014ida,Dima:2017pwp,Crisostomi:2017lbg,Langlois:2017dyl}.

We set $\alpha_{B0}=-2\times 10^{-3}$ and investigate how the evolution of $\delta$ is modified for different sets of parameters $(\alpha_{H0},\beta_{10})$.
We choose the values of the parameters so that the stability condition discussed in Appendix~\ref{app:para-region} is satisfied.
If the parameters are taken so that $\alpha_{H0}+2\beta_{10}\simeq 0$,
the situation is similar to what we have discussed in the case of $\alpha_{H0}+2\beta_{10}=0$:
Unless $\alpha_{H0}$ and $\beta_{10}$ are several orders of magnitude smaller than $\alpha_{B0}$, collapse ceases to proceed at some moment when $x$ becomes imaginary.
We therefore choose $(\alpha_{H0},\beta_{10})$ such that they are sufficiently distant from the line $\alpha_{H0}+2\beta_{10}=0$ in the parameter space.
It is then easy to find the parameters for which collapse is not halted before completion.
The initial condition for $\delta$ is set as Eq.~\eqref{eq:init-delta}.

First, let us discuss the case of the background model with Eq.~\eqref{Gcos1}.
Our numerical results for this background model are shown in Fig.~\ref{collapse_cubic}.
It can be seen that the time of collapse, $t_{\mrm{col}}$, is delayed as the values of $\alpha_{H0}$ and $\beta_{10}$ increase when compared under the same initial conditions for $\delta$.
By comparing the results for two different pairs of the parameters $(\alpha_{H0},\beta_{10})$ with the same value of $\gamma_0$, it can be confirmed that the evolution of $\delta$ is determined almost entirely by the value of $\gamma_0$ governing the early-time linear evolution of $\delta$.
The aforementioned qualitative behavior can therefore be understood as follows:
for larger $\alpha_{H0}$ and $\beta_{10}$, $\gamma_0$ gets smaller and hence gravity is weaker initially (see Eq.~\eqref{eq:linear}), delaying the time of collapse.

Let us next discuss the case of the background model with $\gamma_0=1$,
in which the effects of the modified evolution of $\delta$ in the early-time linear regime are minimized.
Our numerical results for this background model are shown in Fig.~\ref{collapse_cubic2}.
Similar to the background model with $\gamma_0\neq 1$ discussed above, it can be seen that the collapse time is delayed for larger values of $\alpha_{H0}$ and $\beta_{10}$.
However, the magnitude of the corrections is smaller as compared to the case with $\gamma_0\neq 1$ because modification to GR manifests only at late times in the $\gamma_0=1$ background model.
The relative shift of $t_{\mrm{col}}$ is of order $\mcal{O}(\alpha_{H0},\beta_{10})$.

\section{Implications for the halo mass function}
\label{sec:Mass_function}

A dark matter halo forms after an overdense region collapses. As shown in Fig.~\ref{collapse_cubic}, the collapse time depends on the EFT parameter, and this may affect the number density of the dark matter halos. To quantify it, we calculate the halo mass function based on the Press-Schechter formalism \cite{Press_Schechter_1974}.
The differential mass function is given by
\begin{align}\label{mass}
  \frac{dn(\mathcal{M},t)}{d\mathcal M}=\sqrt{\frac{2}{\pi}}\frac{\bar{\rho}(t_0)}{\mathcal M}\frac{\delta_c(t)}{\sigma^2(\mathcal M)}\left|\frac{d\sigma(\mathcal M)}{d\mathcal M}\right|\exp\left[-\frac{\delta_c^2(t)}{2\sigma^2(\mathcal M)}\right],
\end{align}
where $\delta_c$ is the critical density contrast at time $t$, as will be discussed in more detail below, $\bar{\rho}(t_0)$ is the background density given by Eq.~(\ref{eq:background_density})
and the variance is given by
\begin{align}
    \sigma^2(\mathcal M)=\frac{1}{2\pi}\int^{\infty}_{0}dk~k^2W^2(kR)P_L(k,t_0).
\end{align}
The linear power spectrum $P_L(k,t)$ in the background model with $\gamma_0=1$ is computed by using the only Boltzmann solver currently capable of handling DHOST theories~\cite{Hiramatsu:2020fcd,Hiramatsu:2022ahs,Hiramatsu:2022fgn},
and the window function $W(kR)=3(kR)^{-3}(\sin{kR}-kr\cos{kR})$ is the Fourier transform of the top-hat filter with the comoving radius $R$.
The enclosed mass $\mcal{M}$ is related to the comoving radius $R$ and the matter energy density as $\mathcal{M}=4\pi R^3\bar{\rho}(t_0)/3$.

\begin{figure}[tb]
    \centering
    \begin{minipage}[b]{0.45\linewidth}
    \centering
    \includegraphics[keepaspectratio, scale=0.45]{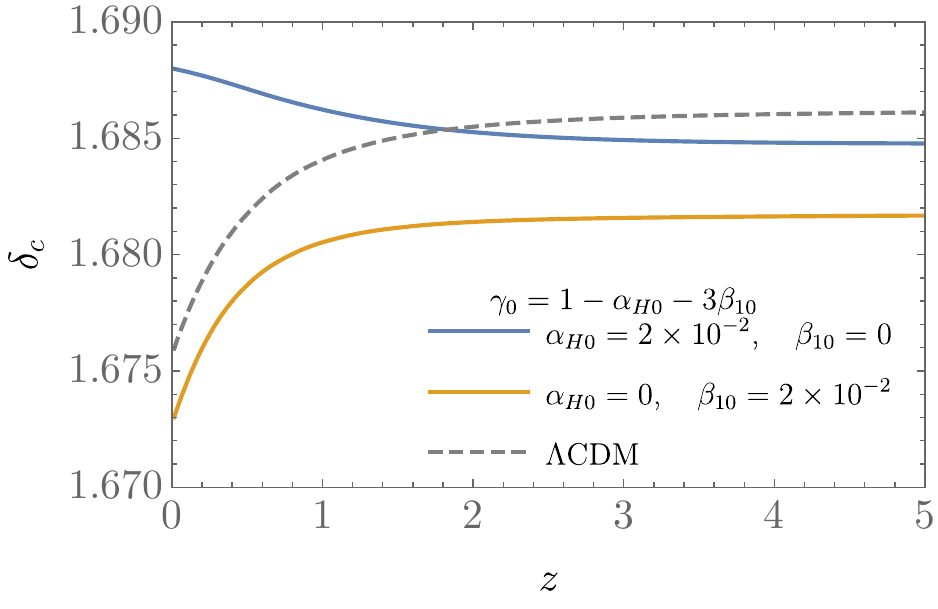}
  \end{minipage} 
  \begin{minipage}[b]{0.45\linewidth}
    \centering
    \includegraphics[keepaspectratio, scale=0.45]{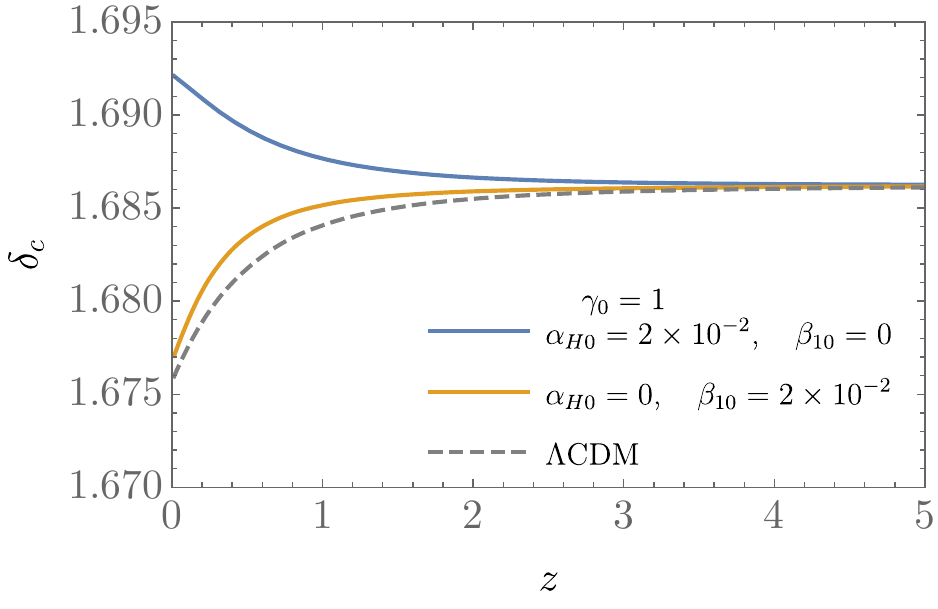}
  \end{minipage} \\
  \caption{Critical density contrast as a function of the redshift for the two background models.
  The parameters are given by $(\alpha_{B0},\alpha_{H0},\beta_{10})=(0,2\times10^{-2},0)$ (solid blue line)
  and $(\alpha_{B0},\alpha_{H0},\beta_{10})=(0,0,2\times10^{-2})$ (solid orange line). The dashed line represents the critical density contrast in the $\Lambda$CDM model.}
 \label{delta_c}
\end{figure}

\begin{figure}[tb]
  \begin{minipage}[b]{0.45\linewidth}
    \centering
    \includegraphics[keepaspectratio, scale=0.45]{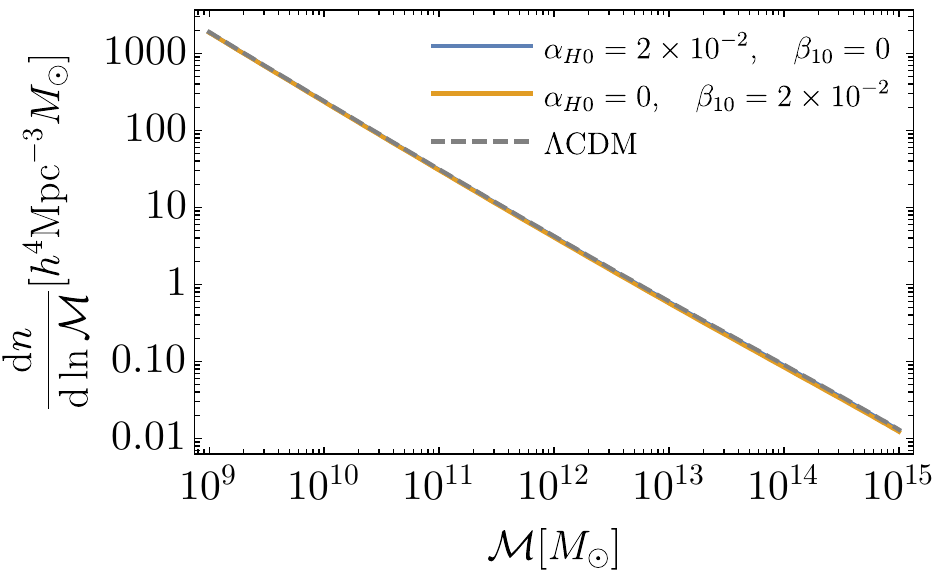}
  \end{minipage} 
  \begin{minipage}[b]{0.45\linewidth}
    \centering
    \includegraphics[keepaspectratio, scale=0.45]{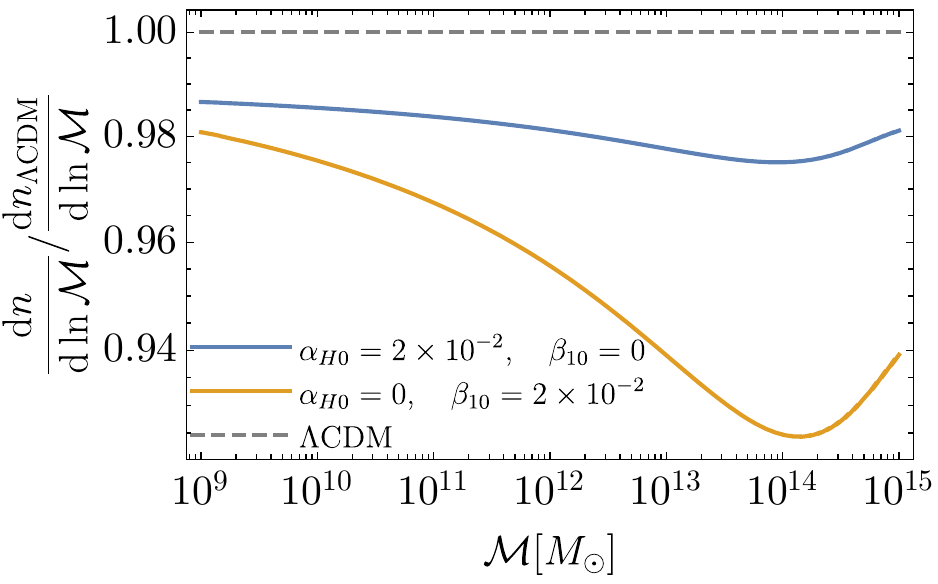}
  \end{minipage} \\
  \caption{Mass function and its deviation from $\Lambda$CDM model. The left panel shows the mass function Eq.~\eqref{mass} at the present time. The parameters are given by $(\alpha_{B0},\alpha_{H0},\beta_{10})=(0,2\times10^{-2},0)$ (solid blue line)
  and $(\alpha_{B0},\alpha_{H0},\beta_{10})=(0,0,2\times10^{-2})$ (solid orange line). The dashed line is the mass function of $\Lambda\mrm{CDM}$ model. The right panel shows the ratio between the mass function obtained from DHOST theory and $\Lambda\mrm{CDM}$ model.}
 \label{mass_function}
\end{figure}

We calculate the critical density contrast $\delta_c$ for selected values of the parameters $(\alpha_{B0},\alpha_{H0},\beta_{10})$ by solving the linearized equation~\eqref{eq:linear} and setting $\delta_c := \delta_L(t_{\rm col})$, where $t_{\mrm{col}}$ is obtained from the spherical collapse model.
The resultant critical density contrast is shown in Fig.~\ref{delta_c} as a function of the redshift $z=a^{-1}(t_{\mrm{col}})-1$ for the background models with $\gamma_0=1-\alpha_{H0}-3\beta_{10}$ and $\gamma_0=1$.
The parameters are chosen as $(\alpha_{B0},\alpha_{H0},\beta_{10}) = (0,2\times 10^{-2},0)$ and $(0,0,2\times 10^{-2})$.
Focusing on the background model with $\gamma_0=1$, we present in Fig.~\ref{mass_function} the halo mass function and its relative deviation from that of the $\Lambda$CDM model evaluated at the present time, showing that the mass function is suppressed compared to that in the $\Lambda$CDM model when $\alpha_{H0}$ and $\beta_{10}$ are taken to be positive.
For $(\alpha_{B0},\alpha_{H0},\beta_{10}) = (0,0,2\times 10^{-2})$, the mass function around $\mcal{M}\sim 10^{14}\,M_\odot$ is reduced by about 10\%.


\section{Conclusions}\label{sec:conclusions}

In this paper, we have studied the spherical collapse model in the effective field theory (EFT) of dark energy, exploring the potential impacts of scalar-tensor theories beyond Horndeski, i.e., degenerate higher-order scalar-tensor (DHOST) theories, on the nonlinear evolution of large-scale structure.
We have not assumed any specific form of the DHOST action, but instead we have assumed the evolution of the background Friedmann universe and the time-dependence of the EFT coefficients.
We have considered two background models.
While both have the same cosmic expansion history as the $\Lambda$CDM model, they are distinguished by the assumption on the ``cosmological'' gravitational constant characterized by a constant parameter denoted as $\gamma_0$.
In the first background model, we have $\gamma_0\neq 1$, leading to the modified evolution of density perturbations already in the linear regime.
The second background model has $\gamma_0=1$. In this case, the modification in the linear regime is minimized, and corrections to the $\Lambda$CDM model manifest mainly at late times.
The EFT functions $\alpha_B$, $\alpha_H$, and $\beta_1$ are assumed to be proportional to
$1-\Omega_{\mrm{m}}(t)$, where $\Omega_{\mrm{m}}(t)$ is the time-dependent matter density parameter, while the other EFT functions are set to zero partially for simplicity and partially to satisfy the constraint on the speed of gravitational waves.

We have carried out separate analyses for the special case where $\alpha_H+2\beta_1=0$ and the generic case where this condition is not satisfied.
On a phenomenological side, this condition implies that gravitons do not decay into the scalar field.
Technically, the algebraic equation that holds for the gradient of the scalar field is quadratic when $\alpha_H+2\beta_1=0$, while it is cubic in the generic case.

In the case where $\alpha_H+2\beta_1=0$, we found that, for $\beta_{10}$ above a certain value, the spherical collapse model fails to describe the entire evolution of the collapsing region, because real solutions for the scalar-field fluctuation cease to exist at some moment.
Here, the subscript 0 stands for the present value of the EFT function.
By requiring that this does not occur until the present time for any initial amplitudes $A_i$,
we have derived a constraint $\beta_{10}< 10^{-7}$ for the fiducial value $\alpha_{B0}=-2\times10^{-3}$.
This constraint is five orders of magnitude stronger than the one obtained from the Hulse-Taylor pulsar~\cite{Hirano:2019scf,Dima:2017pwp}.

In the generic case with $\alpha_H+2\beta_1\neq0$, the above problem is circumvented and the spherical collapse successfully proceeds provided that $(\alpha_H,\beta_1)$ lies sufficiently away from the special line $\alpha_H+2\beta_1=0$.
We investigated how the evolution of the spherical overdense region depends on the ``beyond-Horndeski'' parameters $(\alpha_H,\beta_1)$ in the aforementioned two background models.
In the first background model with $\gamma_0\neq 1$, modification to general relativity arises already in the linear regime due to the deviation of the ``cosmological'' gravitational constant from the Newtonian gravitational constant.
For positive $\alpha_{H0}$ and $\beta_{10}$, we have $\gamma_0<1$, which stunts the growth of density perturbations in the linear regime.
This plays the most significant role in delaying the collapse, while the late-time effects have only a minor impact.
In the second background model with $\gamma_0=1$, density perturbations grow initially in the same way as in the linear regime of the $\Lambda$CDM model.
In this case, the effects of modified gravity arise in the later stage in the nonlinear regime.
We have found that also in this case positive $\alpha_{H0}$ and $\beta_{10}$ delay the collapse, but quantitatively their impacts are smaller than in the $\gamma_0\neq 1$ case when compared at the same $(\alpha_{H0},\beta_{10})$.
As a demonstration, we have computed the halo mass function using the Press-Schechter formalism for selected values of the EFT parameters.
It was found that the mass function is suppressed relative to that in the $\Lambda$CDM model
when $\alpha_{H0}$ and $\beta_{10}$ are taken to be positive.

This paper has formulated the spherical collapse model within the framework of DHOST theories and identified the parameter regions in which an overdense region fails to collapse.
This serves as a first step toward confronting theoretical predictions of DHOST theories with observations of large-scale structure.
It would be interesting to investigate the nonlinear evolution of matter overdensities using $N$-body simulations to confirm whether an overdense region indeed fails to collapse as suggested by the results of the spherical collapse model.
Additionally, it is worth examining whether the Press-Schechter formalism remains valid in DHOST theories by calculating the mass function with $N$-body simulations.

\acknowledgments
We thank Shun Arai for interesting discussions.
The work of TH was supported by
JSPS KAKENHI Grant No.~JP21K03559 and No.~JP23H00110.
The work of TK was supported by
JSPS KAKENHI Grant No.~JP25K07308 and
MEXT-JSPS Grant-in-Aid for Transformative Research Areas (A) ``Extreme Universe'',
No.~JP21H05182 and No.~JP21H05189.

\appendix 
\section{Relation between DHOST theories and EFT of dark energy}
\label{app:EFT}

\subsection{The action for quadratic DHOST theories}

The action for quadratic DHOST theories includes all possible quadratic terms built out of second derivatives of the scalar field $\phi$ and is given by~\cite{Langlois:2015cwa}
\begin{align}
 \label{actionDHOST}
  S=\int d^4x\sqrt{-g}\left[P(\phi,X)+Q(\phi,X)\Box\phi+f(\phi,X){}^{(4)}R+\sum_{I=1}^{5}a_I(\phi,X)L_I(\phi,\phi_{;\nu},\phi_{\rho\sigma})\right],
\end{align}
where $X=-\phi_{;\mu}\phi^{;\mu}/2$, ${}^{(4)}R$ is the four-dimensional Ricci scalar, and $L_I$ $(I=1, \dots, 5)$ are defined as
\begin{align}
  &L_1=\phi_{;\mu\nu}\phi^{\mu\nu},\qquad L_2=(\phi^{;\mu}_{;\mu})^2,\qquad L_3=\phi^{;\mu}_{;\mu}\phi^{;\rho}\phi_{;\rho\sigma}\phi^{;\sigma},
  \notag \\
  &L_4=\phi^{;\mu}\phi_{;\mu\nu}\phi^{;\nu\rho}\phi_{;\rho},\qquad L_5=(\phi^{;\rho}\phi_{;\rho\sigma}\phi^{;\sigma})^2,
\end{align}
with a semicolon denoting the covariant derivative.
In order for the theory to propagate one scalar and two tensor degrees of freedom,
the functions $f$ and $a_I$ must obey degeneracy conditions.
Hereafter, we will focus on the subclass of DHOST theories satisfying the degeneracy condition $a_1+a_2=0$, which is called class Ia.
Other subclasses are not physically interesting because they exhibit instabilities either in the scalar or tensor sector on a cosmological background~\cite{Langlois:2017mxy,deRham:2016wji}.
In class Ia DHOST theories, the remaining degeneracy conditions yield
\begin{align}
    a_4&=-\frac{1}{2f}\left(
        2fa_3-3f_{,X}^2-2Xf_{,X}a_3+X^2a_3^2
    \right),
    \\ 
    a_5&=-\frac{a_3}{f}\left(f_{,X}+Xa_3\right),
\end{align}
leaving two independent function $f$ and $a_3$ in addition to $P$ and $Q$.

\subsection{EFT functions from class Ia DHOST theories}

The $\alpha$-basis representation of the EFT of dark energy is conveniently used in the literature to compare theoretical predictions with observational data.
One can map the DHOST action to the $\alpha$-basis representation of the EFT of dark energy following \cite{Langlois:2017mxy} and~\cite{Dima:2017pwp}.

We write the perturbations around the flat FLRW metric,
$\D s^2=-\D t^2+a^2(t)\delta_{ij}\D x^i\D x^j$,
in the unitary gauge,
using the lapse function $N$, the shift vector $N_i$,
and the three-dimensional metric $h_{ij}$.
The DHOST action is then expressed in terms of $N$, $h$,
the extrinsic and intrinsic curvature tensors of constant $t$ hypersurfaces
$K_{ij}$ and $^{(3)}R_{ij}$, and the first derivatives of the lapse function
$V:=\left(\dot N-N^i\partial_iN\right)/N$ and $a_i:=\partial_i\ln N$.
When expanded around the FLRW metric, the action takes the following form:
\begin{align}
  \label{actionQS}
    S=\int d^4x\sqrt{h}\frac{M^2}{2}\Big[&-(1+\delta N)\delta \mathcal{K}_2+(1+\alpha_T){}^{(3)}R+H^2\alpha_K\delta N^2+4H\alpha_{B}\delta K\delta N+(1+\alpha_{H}){}^{(3)}R\delta N
    \notag \\
  &+4\beta_1\delta KV+\beta_2V^2+\beta_3a_ia^i+\alpha_{V}\delta N\delta\mathcal{K}_2\Big].
\end{align}
where we introduced $H:=\dot{a}/a$, $\delta N:= N-1$, $\delta K^j_i:= K^j_i-H\delta^j_i$, and $\delta \mathcal{K}_2:=\delta K^2-\delta K^j_i\delta K^i_j$,
and dropped the terms that are nonlinear in perturbations but irrelevant to the Vainshtein mechanism.
By a direct calculation, one can relate the coefficient of each term with the functions in the DHOST action as~\cite{Dima:2017pwp}
\begin{align}
  M^2&=2(f-2a_2X),
  \\ 
    \alpha_{B}&=\alpha_{V}-3\beta_1+\frac{\dot{\phi}(f_{,\phi}+2Xf_{,\phi X}+XQ_{,X})}{M^2},\\
    \alpha_{T}&=\frac{2f}{M^2}-1,\\
    \alpha_{H}&=\frac{4X(a_2-f_{,X})}{M^2},\\
    \beta_1&=\frac{2X(f_{,X}-a_2+a_3X)}{M^2},\\
    \beta_2&=-\frac{8X^2(a_3+a_4-2a_5X)}{M^2},\\
    \beta_3&=-\frac{8X(f_{,X}-a_2-a_4X)}{M^2},\\
    \alpha_V&=\frac{4X(f_{,X}-2a_2-2Xa_{2,X})}{M^2}.
\end{align}
The explicit expression for $\alpha_K$ is complicated.
These coefficients are functions of time and the background cosmology determines their time-dependence.
In terms of these coefficients (the EFT functions),
the class Ia degeneracy conditions read 
\begin{align}
    \beta_2=-6\beta_1^2,    \qquad \beta_3=-2\beta_1[2(1+\alpha_H)+\beta_1(1+\alpha_T)].
\end{align}

The physical significance of each coefficient is as follows:
$\alpha_B$ measures the kinetic mixing between the scalar and tensor sectors,
$\alpha_T$ parametrizes the difference between the propagation speeds of gravitational waves and photons,
$\alpha_H$ characterizes the kinetic mixing between the matter and scalar sectors,
$\alpha_V$ signals the modification starting at cubic order in perturbations,
and $\beta_1$, $\beta_2$, and $\beta_3$ are the functions specific to DHOST theories beyond Horndeski.
It is also convenient to introduce another function $\alpha_M$ parametrizing the time dependence of the effective Planck mass~\cite{Bellini:2014fua},
\begin{align}
  \alpha_M:=\frac{\D\ln M^2}{\D\ln a},
\end{align}
though we assume $\alpha_M=0$ for simplicity in the main text.

\section{Coefficients in Eq.~\eqref{xeq}}\label{app:coefficient}

The explicit expression for the coefficients in Eq.~\eqref{xeq} is given by
\begin{align}
            H_0\mcal{C}_2&= \left(\beta _1-1\right) \left(4 \dot{\beta} _1+\dot{\alpha} _H\right)- H \left[\alpha _B \left(4
   \beta _1+\alpha _H-2\right)-2 \left(\beta _1+\alpha _H\right) \left(3 \beta _1+\alpha
   _H\right)+\alpha _H\right],\\
            H_0^2\mcal{C}_1&=H^2
            \biggl\{-\frac{1}{2}\gamma_0 \om \left[\beta _1 \left(3 \beta _1 (\delta +1)+7 \delta +6\right)+\left(6
   \beta _1 (\delta +1)+5 \delta +6\right) \alpha _H+3 (\delta +1) \alpha _H^2+3\right]
   \notag \\ & \quad 
   -
   \left(\alpha _B+1\right) \left(-\beta _1+\alpha _B-\alpha _H\right)\biggr\}
   -2 \dot{\beta}
   _1^2+ \beta _1 \ddot{\beta} _1+ \ddot{\beta} _1-  \dot{H}\left(\alpha _B+1\right)
   \left(\beta _1+\alpha _H+1\right)
   \notag \\ & \quad 
   - H \left[\dot{\alpha} _B\left(\beta _1+\alpha
   _H+1\right)+\dot{\beta} _1 \left(\beta _1-3 \alpha _B+\alpha _H-2\right)-\left(\alpha
   _B+1\right) \dot{\alpha} _H\right]
   - \dot{\beta} _1 \dot{\alpha} _H+ \ddot{\beta} _1 \alpha _H,\\
   H_0^3\mcal{C}_0&=\frac{1}{2}\gamma_0   H^3 \om \left[\left(\beta _1+\alpha _H\right) \left(3 \beta _1+3 \alpha
   _H+4\right)-\alpha _B\right]\delta
   +2 \gamma_0   H \dot{H} \om \left(\beta _1+\alpha _H\right)
   \left(\beta _1+\alpha _H+1\right)\delta
   \notag \\ & \quad 
   +\frac{1}{2}\gamma_0 H^2 \left[ \om \left(\beta _1+\alpha
   _H\right) \left(\beta _1+\alpha _H+1\right)\dot{\delta}+ \om \left(2 \dot{\beta}_1+\dot{\alpha}
   _H\right)\delta +  \left(\beta _1+\alpha _H\right) \left(\beta _1+\alpha _H+1\right)
   \dot{\Omega}_{\mrm{m}}\delta\right],
\end{align}
where $\gamma_0:=1-\alpha_{H0}-3\beta_{10}$.

\section{Acceptable region of the EFT parameters in light of the Vainshtein mechanism and the stability}\label{app:para-region}

In this appendix, we identify the acceptable region of the parameter space in light of the Vainshtein mechanism and the stability in the case of $\alpha_H+2\beta_1=0$ and in the generic case of $\alpha_H+2\beta_1\neq0$.

First, we consider the case of $\alpha_H+2\beta_1=0$ and show the range of $\beta_{1}$ in which the gravitational potentials inside a static matter source are not very different from the standard ones.
With $\alpha_H+2\beta_1=0$, the solution to Eq.~\eqref{xeq} is given by
\begin{align}
\label{xeq:q}
x=\frac{-\mcal{C}_1+\sqrt{\mcal{C}_1^2-4\mcal{C}_0\mcal{C}_2}}{2\mcal{C}_2}.
\end{align}
Here, one of the two roots has been chosen appropriately so that it reduces to the linear solution when $\delta \ll 1$~\cite{Crisostomi:2019yfo}.
In the highly nonlinear regime, $\delta\gg 1$, we have the following two possibilities, depending on whether $\beta_1$ is positive or negative.
In the case of $\beta_1<0$,
Eq.~\eqref{xeq:q} reduces to
\begin{align}
    \label{solx:beta_minus}
    x\simeq-\frac{3\gamma_0H^2\beta_1\om\delta}{4(H\alpha_B+H\beta_1-\dot{\beta}_1)},
\end{align}
while in the case of $\beta_1>0$
we have
\begin{align}
\label{solx:beta_plus}
    x\simeq\frac{H(\alpha_B+\beta_1)}{3H_0\beta_1(1-\beta_1)}.
\end{align}
Substituting Eq.~\eqref{solx:beta_minus} into Eqs.~\eqref{yeqintNL} and \eqref{zeqintNL},
we obtain
\begin{align}
\label{soly:beta_minus}
    y&\simeq\frac{9\gamma_0^2H^4\beta_1^3\om^2\delta^2}{16H_0^2(1-\beta_1)(H\alpha_B+H\beta_1-\dot{\beta}_1)^2},\\
    z&\simeq-\frac{9\gamma_0^2H^4\beta_1^3\om^2\delta^2}{16H_0^2(1-\beta_1)(H\alpha_B+H\beta_1-\dot{\beta}_1)^2},
\end{align}
which is far from the Newtonian behavior and hence is ruled out.
On the other hand, substituting Eq.~\eqref{solx:beta_plus} into Eqs.~\eqref{yeqintNL} and \eqref{zeqintNL}, we find
\begin{align}
\label{soly:beta_plus}
    y&\simeq\frac{\gamma_0H^2\om\delta}{2H_0^2(1-\beta_1)^2},\\
    z&\simeq\frac{\gamma_0H^2(1-2\beta_1)\om\delta}{2H_0^2(1-\beta_1)^2}.
\end{align}
Aside from the deviation of $\mcal{O}(\beta_1)$ due to the partial breaking of Vainshtein screening, the appropriate behavior of the gravitational potentials is reproduced.
Thus, we require that, in the case of $\alpha_H+2\beta_1=0$,
\begin{align}
    \beta_1>0.
\end{align}

Let us then derive the condition on $\alpha_B$ implied by
the stability of the scalar perturbation.
According to Ref.~\cite{Langlois:2017mxy}, the sound speed squared
in the absence of matter is given by
\begin{align}
    c_s^2=-\frac{B_{\tilde\zeta}}{A_{\tilde\zeta}},
\end{align}
where
\begin{align}
    A_{\tilde\zeta}&:=\frac{\alpha}{\left(1+\alpha_B-\dot{\beta}_1/H\right)^2},\\
    B_{\tilde\zeta}&:=2(1+\alpha_T)-\frac{2}{aM^2}\frac{\D}{\D t}\left[\frac{aM^2(1+\alpha_H)+\beta_1(1+\alpha_T)}{H(1+\alpha_B)-\dot{\beta}_1}\right],
\end{align}
with
\begin{align}
    \alpha:=\alpha_K+6\alpha_B^2-\frac{6}{a^3H^2M^3}\frac{\D}{\D t}(a^3HM^2\alpha_B\beta_1).
\end{align}
Here, $\alpha_K$ is defined as the coefficient of $\delta N^2$ in the EFT action.
The expression for $\alpha_K$ in terms of the functions in the DHOST action~\eqref{actionDHOST} is found in Refs.~\cite{Arai:2019zul,Hiramatsu:2020fcd}.  
The ghost instability is avoided if $A_{\tilde\zeta}>0$.
When $\alpha_B$ and $\beta_1$ are small, we have $A_{\tilde \zeta}\simeq \alpha_K$
and hence $\alpha_K>0$.
In the presence of matter, the sound speed is corrected as 
\begin{align}
\label{soundspeed}
    \hat{c}_s^2:=c_s^2-\frac{3\om}{\alpha}\left[1+\alpha_H+\beta_1(1+\alpha_T)\right]^2.
\end{align}
In the matter dominant era, $\hat{c}_s^2$ reduces to
\begin{align}
    \hat{c}_s^2=\frac{5\left(-\alpha_{B0}+2\beta_{10}\right)}{\alpha_{K0}},
\end{align}
where we assumed that
$\alpha_T=\alpha_M=0$, $\alpha_H+2\beta_1=0$, and $\alpha_K$ scales as $\alpha_K=\alpha_{K0}(1-\om)/(1-\omp)$.
Given that $\alpha_{K0}>0$, we obtain the condition 
\begin{align}
    -\alpha_{B0}+2\beta_{10}>0
\end{align}
to avoid the gradient instability.

\begin{figure}[tb]
    \centering
    \includegraphics[keepaspectratio, scale=0.5]{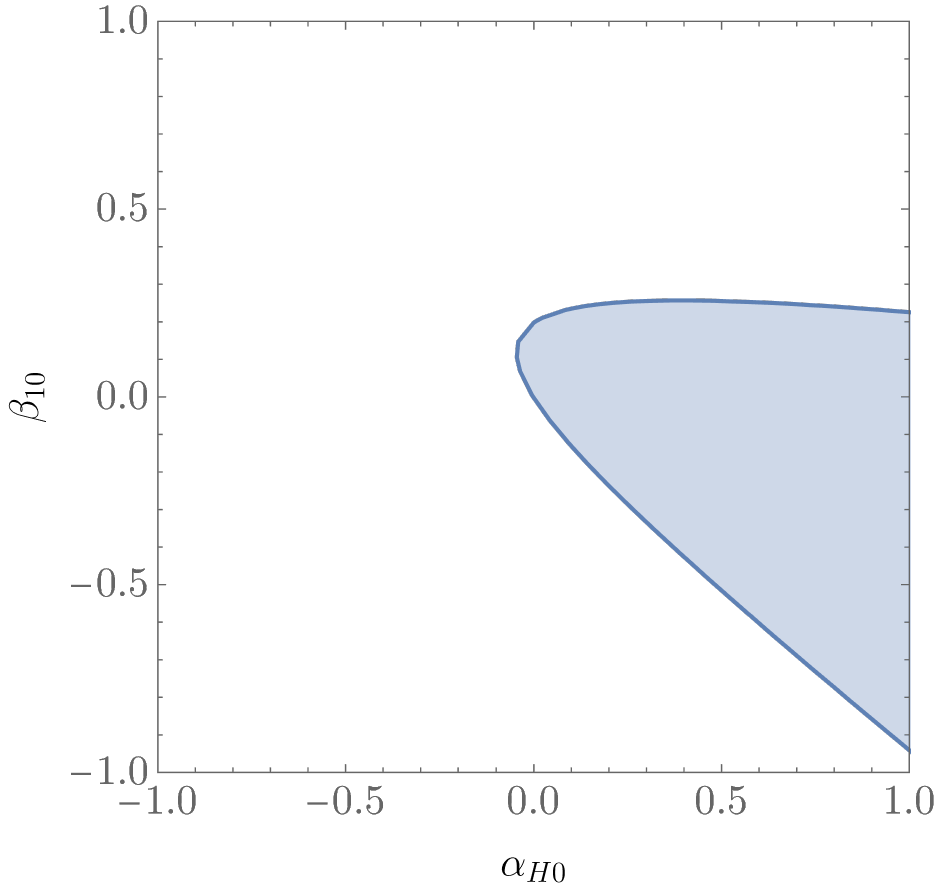}
    \label{Fig:soundspeed}
    \caption{Parameter region satisfying $\hat{c}_s^2>0$.}
\end{figure}

Next, we consider the generic case of $\alpha_H+2\beta_1\neq0$
and present the acceptable parameter region for $(\alpha_{H0},\beta_{10})$ with fixed $\alpha_K$ and $\alpha_B$.
The sound speed squared now becomes
\begin{align}
     \hat{c}_s^2&=\frac{3 \omp}{6 \alpha
   _{B0}^2+\alpha_{K0}-9 \beta _{10} (3
   \omp+2) \alpha _{B0}} \left[-3 \beta _{10}+\alpha
   _{B0} \left(5 \beta _{10}+\alpha
   _{H0}-1\right)-\left(\beta _{10}+\alpha
   _{H0}\right) \left(9 \beta _{10}+\alpha
   _{H0}\right)+\alpha _{H0}\right]
   \notag \\ &\quad 
   -2
   \left(\alpha _{B0}+1\right) \left(-\beta
   _{10}+\alpha _{B0}-\alpha
   _{H0}\right)+9 \beta _{10} \omp^2 \left(-\beta _{10}+\alpha
   _{H0}+3\right).
\end{align}
Taking for instance $\om=\omp=0.3153$, $\alpha_{K0}=1$, and $\alpha_{B0}=-2\times10^{-3}$ yields the colored region in Fig.~\ref{Fig:soundspeed} that satisfies the stability condition $\hat c_s^2>0$.

The above stability argument is only for the early stage
in the matter-dominated era. In the analysis presented in the main text,
we confirm that the stability condition is satisfied not only initially but also throughout
the entire evolution.

\bibliography{refs}

\providecommand{\href}[2]{#2}\begingroup\raggedright\begin{thebibliography}{10}

\bibitem{SupernovaCosmologyProject:1998vns}
{\scshape Supernova Cosmology Project} collaboration, S.~Perlmutter et~al., \emph{{Measurements of $\Omega$ and $\Lambda$ from 42 High Redshift Supernovae}}, \href{https://doi.org/10.1086/307221}{\emph{Astrophys. J.} {\bfseries 517} (1999) 565} [\href{https://arxiv.org/abs/astro-ph/9812133}{{\ttfamily astro-ph/9812133}}].

\bibitem{SupernovaSearchTeam:1998fmf}
{\scshape Supernova Search Team} collaboration, A.~G. Riess et~al., \emph{{Observational evidence from supernovae for an accelerating universe and a cosmological constant}}, \href{https://doi.org/10.1086/300499}{\emph{Astron. J.} {\bfseries 116} (1998) 1009} [\href{https://arxiv.org/abs/astro-ph/9805201}{{\ttfamily astro-ph/9805201}}].

\bibitem{Langlois:2015cwa}
D.~Langlois and K.~Noui, \emph{{Degenerate higher derivative theories beyond Horndeski: evading the Ostrogradski instability}}, \href{https://doi.org/10.1088/1475-7516/2016/02/034}{\emph{JCAP} {\bfseries 02} (2016) 034} [\href{https://arxiv.org/abs/1510.06930}{{\ttfamily 1510.06930}}].

\bibitem{Crisostomi:2016czh}
M.~Crisostomi, K.~Koyama and G.~Tasinato, \emph{{Extended Scalar-Tensor Theories of Gravity}}, \href{https://doi.org/10.1088/1475-7516/2016/04/044}{\emph{JCAP} {\bfseries 04} (2016) 044} [\href{https://arxiv.org/abs/1602.03119}{{\ttfamily 1602.03119}}].

\bibitem{BenAchour:2016fzp}
J.~Ben~Achour, M.~Crisostomi, K.~Koyama, D.~Langlois, K.~Noui and G.~Tasinato, \emph{{Degenerate higher order scalar-tensor theories beyond Horndeski up to cubic order}}, \href{https://doi.org/10.1007/JHEP12(2016)100}{\emph{JHEP} {\bfseries 12} (2016) 100} [\href{https://arxiv.org/abs/1608.08135}{{\ttfamily 1608.08135}}].

\bibitem{Horndeski:1974wa}
G.~W. Horndeski, \emph{{Second-order scalar-tensor field equations in a four-dimensional space}}, \href{https://doi.org/10.1007/BF01807638}{\emph{Int. J. Theor. Phys.} {\bfseries 10} (1974) 363}.

\bibitem{Langlois:2018dxi}
D.~Langlois, \emph{{Dark energy and modified gravity in degenerate higher-order scalar\textendash{}tensor (DHOST) theories: A review}}, \href{https://doi.org/10.1142/S0218271819420069}{\emph{Int. J. Mod. Phys. D} {\bfseries 28} (2019) 1942006} [\href{https://arxiv.org/abs/1811.06271}{{\ttfamily 1811.06271}}].

\bibitem{Kobayashi:2019hrl}
T.~Kobayashi, \emph{{Horndeski theory and beyond: a review}}, \href{https://doi.org/10.1088/1361-6633/ab2429}{\emph{Rept. Prog. Phys.} {\bfseries 82} (2019) 086901} [\href{https://arxiv.org/abs/1901.07183}{{\ttfamily 1901.07183}}].

\bibitem{Crisostomi:2018bsp}
M.~Crisostomi, K.~Koyama, D.~Langlois, K.~Noui and D.~A. Steer, \emph{{Cosmological evolution in DHOST theories}}, \href{https://doi.org/10.1088/1475-7516/2019/01/030}{\emph{JCAP} {\bfseries 01} (2019) 030} [\href{https://arxiv.org/abs/1810.12070}{{\ttfamily 1810.12070}}].

\bibitem{Hirano:2019nkz}
S.~Hirano, T.~Kobayashi, D.~Yamauchi and S.~Yokoyama, \emph{{Constraining degenerate higher-order scalar-tensor theories with linear growth of matter density fluctuations}}, \href{https://doi.org/10.1103/PhysRevD.99.104051}{\emph{Phys. Rev. D} {\bfseries 99} (2019) 104051} [\href{https://arxiv.org/abs/1902.02946}{{\ttfamily 1902.02946}}].

\bibitem{Hiramatsu:2020fcd}
T.~Hiramatsu and D.~Yamauchi, \emph{{Testing gravity theories with cosmic microwave background in the degenerate higher-order scalar-tensor theory}}, \href{https://doi.org/10.1103/PhysRevD.102.083525}{\emph{Phys. Rev. D} {\bfseries 102} (2020) 083525} [\href{https://arxiv.org/abs/2004.09520}{{\ttfamily 2004.09520}}].

\bibitem{Hiramatsu:2022fgn}
T.~Hiramatsu, \emph{{CMB constraints on DHOST theories}}, \href{https://doi.org/10.1088/1475-7516/2022/10/035}{\emph{JCAP} {\bfseries 10} (2022) 035} [\href{https://arxiv.org/abs/2205.11559}{{\ttfamily 2205.11559}}].

\bibitem{Brax:2025osk}
P.~Brax and A.~Lazanu, \emph{{Primordial gravitational waves in DHOST inflation}},  \href{https://arxiv.org/abs/2501.13210}{{\ttfamily 2501.13210}}.

\bibitem{Kobayashi:2014ida}
T.~Kobayashi, Y.~Watanabe and D.~Yamauchi, \emph{{Breaking of Vainshtein screening in scalar-tensor theories beyond Horndeski}}, \href{https://doi.org/10.1103/PhysRevD.91.064013}{\emph{Phys. Rev. D} {\bfseries 91} (2015) 064013} [\href{https://arxiv.org/abs/1411.4130}{{\ttfamily 1411.4130}}].

\bibitem{Langlois:2017dyl}
D.~Langlois, R.~Saito, D.~Yamauchi and K.~Noui, \emph{{Scalar-tensor theories and modified gravity in the wake of GW170817}}, \href{https://doi.org/10.1103/PhysRevD.97.061501}{\emph{Phys. Rev. D} {\bfseries 97} (2018) 061501} [\href{https://arxiv.org/abs/1711.07403}{{\ttfamily 1711.07403}}].

\bibitem{Crisostomi:2017lbg}
M.~Crisostomi and K.~Koyama, \emph{{Vainshtein mechanism after GW170817}}, \href{https://doi.org/10.1103/PhysRevD.97.021301}{\emph{Phys. Rev. D} {\bfseries 97} (2018) 021301} [\href{https://arxiv.org/abs/1711.06661}{{\ttfamily 1711.06661}}].

\bibitem{Dima:2017pwp}
A.~Dima and F.~Vernizzi, \emph{{Vainshtein Screening in Scalar-Tensor Theories before and after GW170817: Constraints on Theories beyond Horndeski}}, \href{https://doi.org/10.1103/PhysRevD.97.101302}{\emph{Phys. Rev. D} {\bfseries 97} (2018) 101302} [\href{https://arxiv.org/abs/1712.04731}{{\ttfamily 1712.04731}}].

\bibitem{Hirano:2019scf}
S.~Hirano, T.~Kobayashi and D.~Yamauchi, \emph{{Screening mechanism in degenerate higher-order scalar-tensor theories evading gravitational wave constraints}}, \href{https://doi.org/10.1103/PhysRevD.99.104073}{\emph{Phys. Rev. D} {\bfseries 99} (2019) 104073} [\href{https://arxiv.org/abs/1903.08399}{{\ttfamily 1903.08399}}].

\bibitem{Babichev:2016jom}
E.~Babichev, K.~Koyama, D.~Langlois, R.~Saito and J.~Sakstein, \emph{{Relativistic Stars in Beyond Horndeski Theories}}, \href{https://doi.org/10.1088/0264-9381/33/23/235014}{\emph{Class. Quant. Grav.} {\bfseries 33} (2016) 235014} [\href{https://arxiv.org/abs/1606.06627}{{\ttfamily 1606.06627}}].

\bibitem{Sakstein:2016oel}
J.~Sakstein, E.~Babichev, K.~Koyama, D.~Langlois and R.~Saito, \emph{{Towards Strong Field Tests of Beyond Horndeski Gravity Theories}}, \href{https://doi.org/10.1103/PhysRevD.95.064013}{\emph{Phys. Rev. D} {\bfseries 95} (2017) 064013} [\href{https://arxiv.org/abs/1612.04263}{{\ttfamily 1612.04263}}].

\bibitem{Kobayashi:2018xvr}
T.~Kobayashi and T.~Hiramatsu, \emph{{Relativistic stars in degenerate higher-order scalar-tensor theories after GW170817}}, \href{https://doi.org/10.1103/PhysRevD.97.104012}{\emph{Phys. Rev. D} {\bfseries 97} (2018) 104012} [\href{https://arxiv.org/abs/1803.10510}{{\ttfamily 1803.10510}}].

\bibitem{Kobayashi:2025bdh}
T.~Kobayashi, \emph{{Gravitomagnetic tidal response of relativistic stars in partially screened scalar-tensor theories}},  \href{https://arxiv.org/abs/2501.10659}{{\ttfamily 2501.10659}}.

\bibitem{Kimura:2011dc}
R.~Kimura, T.~Kobayashi and K.~Yamamoto, \emph{{Vainshtein screening in a cosmological background in the most general second-order scalar-tensor theory}}, \href{https://doi.org/10.1103/PhysRevD.85.024023}{\emph{Phys. Rev. D} {\bfseries 85} (2012) 024023} [\href{https://arxiv.org/abs/1111.6749}{{\ttfamily 1111.6749}}].

\bibitem{Narikawa:2013pjr}
T.~Narikawa, T.~Kobayashi, D.~Yamauchi and R.~Saito, \emph{{Testing general scalar-tensor gravity and massive gravity with cluster lensing}}, \href{https://doi.org/10.1103/PhysRevD.87.124006}{\emph{Phys. Rev. D} {\bfseries 87} (2013) 124006} [\href{https://arxiv.org/abs/1302.2311}{{\ttfamily 1302.2311}}].

\bibitem{Koyama:2013paa}
K.~Koyama, G.~Niz and G.~Tasinato, \emph{{Effective theory for the Vainshtein mechanism from the Horndeski action}}, \href{https://doi.org/10.1103/PhysRevD.88.021502}{\emph{Phys. Rev. D} {\bfseries 88} (2013) 021502} [\href{https://arxiv.org/abs/1305.0279}{{\ttfamily 1305.0279}}].

\bibitem{Arai:2022zzz}
S.~Arai, K.~Aoki, Y.~Chinone, R.~Kimura, T.~Kobayashi, H.~Miyatake et~al., \emph{{Cosmological gravity probes: Connecting recent theoretical developments to forthcoming observations}}, \href{https://doi.org/10.1093/ptep/ptad052}{\emph{PTEP} {\bfseries 2023} (2023) 072E01} [\href{https://arxiv.org/abs/2212.09094}{{\ttfamily 2212.09094}}].

\bibitem{Bellini:2012qn}
E.~Bellini, N.~Bartolo and S.~Matarrese, \emph{{Spherical Collapse in covariant Galileon theory}}, \href{https://doi.org/10.1088/1475-7516/2012/06/019}{\emph{JCAP} {\bfseries 06} (2012) 019} [\href{https://arxiv.org/abs/1202.2712}{{\ttfamily 1202.2712}}].

\bibitem{Deffayet:2009mn}
C.~Deffayet, S.~Deser and G.~Esposito-Farese, \emph{{Generalized Galileons: All scalar models whose curved background extensions maintain second-order field equations and stress-tensors}}, \href{https://doi.org/10.1103/PhysRevD.80.064015}{\emph{Phys. Rev. D} {\bfseries 80} (2009) 064015} [\href{https://arxiv.org/abs/0906.1967}{{\ttfamily 0906.1967}}].

\bibitem{Deffayet:2009wt}
C.~Deffayet, G.~Esposito-Farese and A.~Vikman, \emph{{Covariant Galileon}}, \href{https://doi.org/10.1103/PhysRevD.79.084003}{\emph{Phys. Rev. D} {\bfseries 79} (2009) 084003} [\href{https://arxiv.org/abs/0901.1314}{{\ttfamily 0901.1314}}].

\bibitem{Barreira:2013xea}
A.~Barreira, B.~Li, C.~M. Baugh and S.~Pascoli, \emph{{Spherical collapse in Galileon gravity: fifth force solutions, halo mass function and halo bias}}, \href{https://doi.org/10.1088/1475-7516/2013/11/056}{\emph{JCAP} {\bfseries 11} (2013) 056} [\href{https://arxiv.org/abs/1308.3699}{{\ttfamily 1308.3699}}].

\bibitem{Frusciante:2020zfs}
N.~Frusciante and F.~Pace, \emph{{Growth of non-linear structures and spherical collapse in the Galileon Ghost Condensate model}}, \href{https://doi.org/10.1016/j.dark.2020.100686}{\emph{Phys. Dark Univ.} {\bfseries 30} (2020) 100686} [\href{https://arxiv.org/abs/2004.11881}{{\ttfamily 2004.11881}}].

\bibitem{Deffayet:2010qz}
C.~Deffayet, O.~Pujolas, I.~Sawicki and A.~Vikman, \emph{{Imperfect Dark Energy from Kinetic Gravity Braiding}}, \href{https://doi.org/10.1088/1475-7516/2010/10/026}{\emph{JCAP} {\bfseries 10} (2010) 026} [\href{https://arxiv.org/abs/1008.0048}{{\ttfamily 1008.0048}}].

\bibitem{Kase:2018iwp}
R.~Kase and S.~Tsujikawa, \emph{{Dark energy scenario consistent with GW170817 in theories beyond Horndeski gravity}}, \href{https://doi.org/10.1103/PhysRevD.97.103501}{\emph{Phys. Rev. D} {\bfseries 97} (2018) 103501} [\href{https://arxiv.org/abs/1802.02728}{{\ttfamily 1802.02728}}].

\bibitem{Albuquerque:2024hwv}
I.~S. Albuquerque, N.~Frusciante, F.~Pace and C.~Schimd, \emph{{Spherical collapse and halo abundance in shift-symmetric Galileon theory}}, \href{https://doi.org/10.1103/PhysRevD.109.023535}{\emph{Phys. Rev. D} {\bfseries 109} (2024) 023535} [\href{https://arxiv.org/abs/2401.04096}{{\ttfamily 2401.04096}}].

\bibitem{LIGOScientific:2017zic}
{\scshape LIGO Scientific, Virgo, Fermi-GBM, INTEGRAL} collaboration, B.~P. Abbott et~al., \emph{{Gravitational Waves and Gamma-rays from a Binary Neutron Star Merger: GW170817 and GRB 170817A}}, \href{https://doi.org/10.3847/2041-8213/aa920c}{\emph{Astrophys. J. Lett.} {\bfseries 848} (2017) L13} [\href{https://arxiv.org/abs/1710.05834}{{\ttfamily 1710.05834}}].

\bibitem{Creminelli:2017sry}
P.~Creminelli and F.~Vernizzi, \emph{{Dark Energy after GW170817 and GRB170817A}}, \href{https://doi.org/10.1103/PhysRevLett.119.251302}{\emph{Phys. Rev. Lett.} {\bfseries 119} (2017) 251302} [\href{https://arxiv.org/abs/1710.05877}{{\ttfamily 1710.05877}}].

\bibitem{Baker:2017hug}
T.~Baker, E.~Bellini, P.~G. Ferreira, M.~Lagos, J.~Noller and I.~Sawicki, \emph{{Strong constraints on cosmological gravity from GW170817 and GRB 170817A}}, \href{https://doi.org/10.1103/PhysRevLett.119.251301}{\emph{Phys. Rev. Lett.} {\bfseries 119} (2017) 251301} [\href{https://arxiv.org/abs/1710.06394}{{\ttfamily 1710.06394}}].

\bibitem{Sakstein:2017xjx}
J.~Sakstein and B.~Jain, \emph{{Implications of the Neutron Star Merger GW170817 for Cosmological Scalar-Tensor Theories}}, \href{https://doi.org/10.1103/PhysRevLett.119.251303}{\emph{Phys. Rev. Lett.} {\bfseries 119} (2017) 251303} [\href{https://arxiv.org/abs/1710.05893}{{\ttfamily 1710.05893}}].

\bibitem{Ezquiaga:2017ekz}
J.~M. Ezquiaga and M.~Zumalac\'arregui, \emph{{Dark Energy After GW170817: Dead Ends and the Road Ahead}}, \href{https://doi.org/10.1103/PhysRevLett.119.251304}{\emph{Phys. Rev. Lett.} {\bfseries 119} (2017) 251304} [\href{https://arxiv.org/abs/1710.05901}{{\ttfamily 1710.05901}}].

\bibitem{Creminelli:2018xsv}
P.~Creminelli, M.~Lewandowski, G.~Tambalo and F.~Vernizzi, \emph{{Gravitational Wave Decay into Dark Energy}}, \href{https://doi.org/10.1088/1475-7516/2018/12/025}{\emph{JCAP} {\bfseries 12} (2018) 025} [\href{https://arxiv.org/abs/1809.03484}{{\ttfamily 1809.03484}}].

\bibitem{Crisostomi:2019yfo}
M.~Crisostomi, M.~Lewandowski and F.~Vernizzi, \emph{{Vainshtein regime in scalar-tensor gravity: Constraints on degenerate higher-order scalar-tensor theories}}, \href{https://doi.org/10.1103/PhysRevD.100.024025}{\emph{Phys. Rev. D} {\bfseries 100} (2019) 024025} [\href{https://arxiv.org/abs/1903.11591}{{\ttfamily 1903.11591}}].

\bibitem{Press_Schechter_1974}
W.~H. {Press} and P.~{Schechter}, \emph{{Formation of Galaxies and Clusters of Galaxies by Self-Similar Gravitational Condensation}}, \href{https://doi.org/10.1086/152650}{\emph{\apj} {\bfseries 187} (1974) 425}.

\bibitem{Hiramatsu:2022ahs}
T.~Hiramatsu and T.~Kobayashi, \emph{{Testing gravity with the cosmic microwave background: constraints on modified gravity with two tensorial degrees of freedom}}, \href{https://doi.org/10.1088/1475-7516/2022/07/040}{\emph{JCAP} {\bfseries 07} (2022) 040} [\href{https://arxiv.org/abs/2205.04688}{{\ttfamily 2205.04688}}].

\bibitem{Langlois:2017mxy}
D.~Langlois, M.~Mancarella, K.~Noui and F.~Vernizzi, \emph{{Effective Description of Higher-Order Scalar-Tensor Theories}}, \href{https://doi.org/10.1088/1475-7516/2017/05/033}{\emph{JCAP} {\bfseries 05} (2017) 033} [\href{https://arxiv.org/abs/1703.03797}{{\ttfamily 1703.03797}}].

\bibitem{deRham:2016wji}
C.~de~Rham and A.~Matas, \emph{{Ostrogradsky in Theories with Multiple Fields}}, \href{https://doi.org/10.1088/1475-7516/2016/06/041}{\emph{JCAP} {\bfseries 06} (2016) 041} [\href{https://arxiv.org/abs/1604.08638}{{\ttfamily 1604.08638}}].

\bibitem{Bellini:2014fua}
E.~Bellini and I.~Sawicki, \emph{{Maximal freedom at minimum cost: linear large-scale structure in general modifications of gravity}}, \href{https://doi.org/10.1088/1475-7516/2014/07/050}{\emph{JCAP} {\bfseries 07} (2014) 050} [\href{https://arxiv.org/abs/1404.3713}{{\ttfamily 1404.3713}}].

\bibitem{Arai:2019zul}
S.~Arai, P.~Karmakar and A.~Nishizawa, \emph{{Cosmological evolution of viable models in the generalized scalar-tensor theory}}, \href{https://doi.org/10.1103/PhysRevD.102.024003}{\emph{Phys. Rev. D} {\bfseries 102} (2020) 024003} [\href{https://arxiv.org/abs/1912.01768}{{\ttfamily 1912.01768}}].

\end{thebibliography}\endgroup
\bibliographystyle{JHEP}
\end{document}